\documentclass[12pt,a4paper]{article}
\pdfoutput=1

\usepackage{graphicx,array}
\usepackage{color}
\usepackage{latexsym}
\usepackage{amsthm}
\usepackage{amsmath}
\usepackage{amssymb}
\usepackage[table]{colortbl}
\usepackage[utf8]{inputenc}

\usepackage{dsfont}
\usepackage{epsfig}
\usepackage{slashed}
\usepackage{bbold}
\usepackage{psfrag}
\usepackage[svgnames]{xcolor}
\PassOptionsToPackage{caption=false}{subfig}
\usepackage{subcaption}
\usepackage{xfrac}
\usepackage{multirow}
\usepackage{booktabs}
\usepackage[colorlinks=true,linkcolor=Black,citecolor=Green,urlcolor=blue]{hyperref}
\usepackage{cite}
\usepackage[normalem]{ulem}
\usepackage{float} 
\usepackage[bottom]{footmisc} 

\def\bal#1\eal{\begin{align}#1\end{align}}

\def\bel#1{\begin{equation} \label{#1}}

\newcommand{\be}{\begin{equation}}
\newcommand{\ee}{\end{equation}}
\newcommand{\bea}{\begin{eqnarray}}
\newcommand{\eea}{\end{eqnarray}}

\newcommand{\LL}{\mathcal{L}}
\newcommand{\OO}{\mathcal{O}}

\newcommand{\GeV}{\textrm{ GeV}}
\newcommand{\TeV}{\textrm{ TeV}}
\newcommand{\SU}{\textrm{SU}}

\newcommand{\U}{\textrm{U}}

\newcommand{\SM}{\textrm{SM}}

\newcommand{\Br}{\textrm{Br}}

\renewcommand{\Re}{\textrm{Re}}
\renewcommand{\Im}{\textrm{Im}}

\newcommand{\Tab}[1]{Tab.~\ref{tab:#1}}

\newcommand{\Fig}[1]{Fig.~\ref{fig:#1}}

\newcommand{\Eq}[1]{Eq.~(\ref{eq:#1})}

\newcommand{\ord}[1]{\mathcal{O}(#1)}

\usepackage{mathrsfs}

\def\({\left(}
\def\){\right)}

\usepackage{soul}

\begin{document}
\begin{flushright}
\hspace{3cm} 
SISSA 10/2019/FISI
\end{flushright}
\vspace{.6cm}
\begin{center}

\hspace{-0.4cm}

{\LARGE \bf Rank-One Flavor Violation\\\vspace{0.5cm} and $B$-meson anomalies}\vspace{0.5cm}

\vspace{1cm}{Valerio Gherardi$^{a,b}$, David Marzocca$^{b}$, Marco Nardecchia$^{b,c}$, and Andrea Romanino$^{a,b}$}
\\[7mm]
 {\it \small

$^a$ SISSA International School for Advanced Studies, Via Bonomea 265, 34136, Trieste, Italy\\[0.15cm]
$^b$ INFN - Sezione di Trieste, Via Bonomea 265, 34136, Trieste, Italy\\[0.1cm]
$^c$ Sapienza - Universit\'a di Roma, Piazzale Aldo Moro 2, 00185, Roma, Italy\\[0.1cm]
 }

\end{center}

\bigskip \bigskip \bigskip

\centerline{\bf Abstract} 
\begin{quote}
We assume that the quark-flavor coefficients matrix of the semileptonic operators addressing the neutral-current $B$-meson anomalies has rank-one, i.e. it can be described by a single vector in quark-flavor space. By correlating the observed anomalies to other flavor and high-$p_T$ observables, we constrain its possible directions and we show that a large region of the parameter space of this framework will be explored by flavor data from the NA62, KOTO, LHCb and Belle II experiments.
\end{quote}

\newpage
\tableofcontents


\section{Introduction}

The observed deviations from the Standard Model (SM) in $b \to s \mu \mu$ transitions are one of the few pieces of data hinting at the presence of New Physics (NP) at, or near, the TeV scale.
What makes them particularly intriguing is the number of independent observables in which they have been measured and the fact that all deviations can be consistently described by a single NP effect.
The most relevant observables are: Lepton Flavor Universality (LFU) ratios $R_K$ \cite{Aaij:2014ora,Aaij:2019aa} and $R_{K^{*}}$ \cite{Aaij:2017vbb,2019:BelleRKst}, differential branching ratios \cite{Aaij:2014pli} in $b \to s \mu\mu$ transitions, angular distributions in $B \to K^* \mu^+ \mu^-$ \cite{Aaij:2015esa,Aaij:2013qta,Aaij:2015oid}, and the leptonic meson decay $B_s^0 \to \mu^+ \mu^-$ \cite{Aaij:2017vad,Aaboud:2018mst}. The experimental results for the observables with clean SM prediction are collected in Table~\ref{tab:bsmumuObs}.

Model-independent analyses of neutral-current anomalies hint towards NP coupling to quark and lepton vectorial currents \cite{Hiller:2014yaa,Descotes-Genon:2015uva,Altmannshofer:2017fio,Capdevila:2017bsm,DAmico:2017mtc,Altmannshofer:2017yso,Geng:2017svp,Ciuchini:2017mik,Hiller:2017bzc,Alok:2017sui,Hurth:2017hxg,Alguero:2019aa,Alok:2019ufo,Ciuchini:2019usw,Aebischer:2019mlg}.
As a matter of fact, the vast majority of NP explanations of the anomalies boils down, at low energy, to one of the following muonic operators:
\begin{align}
\mathcal{O}_{L} & =(\overline{s}\gamma_{\rho}P_{L}b)(\overline{\mu}\gamma^{\rho}P_{L}\mu)\; , & \mathcal{O}_{9} & =(\overline{s}\gamma_{\rho}P_{L}b)(\overline{\mu}\gamma^{\rho}\mu) \;,
\label{eq:O_L and O_9}
\end{align}
although it has recently been pointed out that allowing for NP in both muons and electrons provides a slight improvement in the fits \cite{Alguero:2018nvb,Alguero:2019pjc,Datta:2019zca}.  In App.~\ref{app:bsmumu} we report a simplified fit of the observables listed in Table~\ref{tab:bsmumuObs}. We also allow for non-vanishing imaginary parts of the operators' coefficients.

\begin{table}[t]
\centering
\begin{tabular}{| c | c | c |} \hline
$R_K~ [1.1, ~6] \GeV^2$ & $0.846 \pm 0.062$ & LHCb \cite{Aaij:2014ora,Aaij:2019aa} \\\hline
\multirow{2}{*}{$R_{K^{*}}~ [0.045, ~1.1] \GeV^2$}  & $0.66 \pm 0.11$ & LHCb \cite{Aaij:2017vbb} \\
		& $0.52^{+ 0.36}_{-0.26}$ & Belle \cite{2019:BelleRKst} \\\hline
\multirow{2}{*}{$R_{K^{*}}~ [1.1, ~6] \GeV^2$}   & $0.69 \pm 0.12$ & LHCb \cite{Aaij:2017vbb} \\
		& $0.96^{+ 0.45}_{-0.29}$ & Belle \cite{2019:BelleRKst} \\\hline
$R_{K^{*}}~ [15, ~19] \GeV^2$ & $1.18^{+ 0.52}_{-0.32}$ & Belle \cite{2019:BelleRKst} \\\hline
\multirow{2}{*}{$\Br(B_s^0 \to \mu\mu) $} & $(3.0^{+0.67}_{-0.63}) \times 10^{-9}$ & LHCb \cite{Aaij:2017vad} \\
 & $(2.8^{+0.8}_{-0.7}) \times 10^{-9}$ & ATLAS \cite{Aaboud:2018mst} \\
\hline
\end{tabular}
\caption{\label{tab:bsmumuObs} Clean observables sensitive to $bs\mu\mu$ contact interactions.}
\end{table}

The two low-energy operators in Eq.~\eqref{eq:O_L and O_9} can be thought to be part of an effective lagrangian involving all the three quark families
\begin{equation}
\mathcal{L}^\text{EFT}_\text{NP} = C^{ij}_L (\overline{d_i}\gamma_{\rho}P_{L}d_j)(\overline{\mu}\gamma^{\rho}P_{L}\mu) +C^{ij}_R (\overline{d_i}\gamma_{\rho}P_{L}d_j)(\overline{\mu}\gamma^{\rho}P_R\mu) \;,
\label{eq:NPEFT}
\end{equation}
where the coefficient of the $\mathcal{O}_L$ operator is identified with $C^{sb}_L$, the coefficient of the $\mathcal{O}_9$ operator with $C^{sb}_L + C^{sb}_R$, and we have focussed on muon processes on the leptonic side. If the $B$-meson anomalies are confirmed by future data, determining the flavor structure of these operators will be crucial for a deeper understanding of the SM flavor puzzle.

Assuming that the relevant NP degrees of freedom lie above the electroweak scale, the natural framework for model-independent studies of the anomalies is actually that of the Standard Model Effective Field Theory (SMEFT). The SMEFT operators that can contribute to the above low-energy ones at the tree level are collected in the following lagrangian:
\be
	\LL ^{\rm SMEFT} _{\rm NP} = C_S^{ij} \left(  \overline{q_{iL}} \gamma_\mu q_{jL} \right)\left( \overline{\ell_{2L}} \gamma^\mu \ell_{2L} \right) 
	+ C_T^{ij} \left(  \overline{q_{iL}} \gamma_\mu \sigma^a q_{jL} \right)\left(  \overline{\ell_{2L}} \gamma^\mu \sigma^a \ell_{2L} \right)
	+ C_R^{ij} \left( \overline{q_{iL}} \gamma_\mu q_{jL} \right)\left( \overline{\mu_R} \gamma^\mu \mu_R \right)  ,
	\label{eq:SMEFTlagrangian}
\ee
giving $C^{ij}_L = C_S^{ij} + C_T^{ij}$ in \Eq{NPEFT}. In the previous equation, $\ell^i_L = \left( \nu_L^i, e_L^i \right)^t$ and $q^i_L = \left( V^*_{ji} u_L^j, d_L^i\right)^t$ are the lepton and quark doublets, in the charged-lepton and down quarks mass basis respectively, and $V$ is the CKM matrix. 

In the above general description, each $d_i \leftrightarrow d_j$ transition corresponds to an independent Wilson coefficient $C^{ij}$, and the experimental data constrain each of them independently. It is however often the case that the underlying new physics gives rise to correlations among the different $C^{ij}$ coefficients. It is also theoretically motivated to expect that this flavor structure is somehow related to the SM Yukawas.
For example, this happens in Minimal Flavor Violation (MFV) \cite{DAmbrosio:2002vsn} and in approaches based on spontaneously broken $\text{U}(2)^n$ flavor symmetries \cite{Barbieri:2011ci,Barbieri:2012uh} (see Refs. \cite{Greljo:2015mma,Barbieri:2015yvd,Barbieri:2016las,Bordone:2017anc,Bordone:2017lsy,Buttazzo:2017ixm,Barbieri:2017tuq} for the link with $B$-anomalies).

In this paper, we consider a different type of correlation. Our key assumption is that the NP sector responsible of the $R_{K^{(*)}}$ signal couples to a single direction in the quark flavor space (as mentioned, we focus here on muon processes on the leptonic side), which requires the Wilson coefficient matrices $C^{ij}_{S,T,R}$ in Eq.~\eqref{eq:SMEFTlagrangian} (and consequently $C^{ij}_{L,R}$ in \Eq{NPEFT}) to be rank-one and proportional:
\be
	C_{S,T,R,L}^{ij} = C_{S,T,R,L} ~ \hat{n}_{i}\hat{n}_{j}^{*}~,
	\label{eq:WCmatrix parametrization}
\ee
where $C_{S,T,R,L} \in \mathbb{R}$, $C_L = C_S + C_T$, and $\hat{n}_{i}$ is a unitary vector in $\U(3)_q$ flavor space.
We dub this scenario \emph{Rank-One Flavor Violation} (ROFV). Rather than being an assumption on the flavor symmetry and its breaking terms (such as Minimal Flavor Violation \cite{DAmbrosio:2002vsn}, for example), this is an assumption on the dynamics underlying these semileptonic operators. 
We refer to \cite{Glashow:2014iga,Bhattacharya:2014wla} for similar approaches in different contexts.

It is perhaps worth emphasizing that our analysis does not rely upon any particular assumption concerning NP effects in the $\tau$ sector, as far as observables with muons are concerned.
Such effects could become relevant only when considering observables with neutrinos, whose flavor is not observed, or loop-generated ones such as $\Delta F=2$ processes in leptoquark models.\footnote{Since experimental limits on semi-tauonic operators are much weaker than those on semi-muonic ones, couplings of new physics to tau leptons can be much larger than to muons, which is consistent with theoretical expectations from NP coupled preferentially to the third family, and for example allows combined explanations of both neutral and charged-current $B$-meson anomalies, see e.g. the analysis in \cite{Buttazzo:2017ixm}.}
On the other hand, we do assume negligible NP effects in the electron sector. This is, by itself, a reasonable assumption since it is supported by data and it is also well motivated in scenarios where NP couplings to leptons follows the same hierarchy as SM Yukawas (such as SU(2)$^5$ flavor symmetries or partial compositeness).

The ROFV assumption is well motivated.
For example, it is automatically realised in all single leptoquark models generating the operators in Eq.~\eqref{eq:O_L and O_9} at low energy\footnote{To be precise, the correlations discussed in the present work apply to all single leptoquark models in which the coupling to electrons is suppressed with respect to the one to muons.}
(see e.g. Ref.~\cite{Angelescu:2018tyl} for a recent comprehensive analysis).
Furthermore, Eq.~\eqref{eq:WCmatrix parametrization} is automatically satisfied in all cases where a single linear combination of SM quark doublets couples to NP:

\be
	\LL \supset \lambda_i \overline{q_{iL}} \OO_{\rm NP} + \text{h.c.}~,
	\label{eq:linearNPlagrangian}
\ee
where $\OO_{\rm NP}$ is an operator involving some beyond the SM degrees of freedom. This condition is actually stronger than strictly required by ROFV, since not only semimuonic operators have rank-one coefficients, but all operators involving quark doublets. This scenario finds realization in several UV models, such as models with single vector-like fermion mediators, and one-loop models with linear flavor violation \cite{Gripaios:2015gra}.
Contrary to the MFV or the minimally broken $\U(2)^5$ scenarios, which predict the flavor structure of all NP contributions, the ROFV assumption is specific to the set of semimuonic operators in Eq.~\eqref{eq:SMEFTlagrangian}. On the other hand, while those scenarios require strong assumptions on the flavor symmetry and its symmetry-breaking terms, ROFV can be accidentally realised from the underlying dynamics, see e.g. Refs.~\cite{Cline:2017aed,Cline:2017qqu}.

From a theoretical point of view it might be natural to expect the direction of the unitary vector $\hat n$ to be close to the third generation of SM quarks. This case is studied in more detail in Sec.~\ref{sec:FlavorSymmetry}. In the following, instead, we abandon any theory prejudice on $\hat n$ and study what are the experimental constraints on its possible directions.
We parametrize $\hat n$ as
\begin{equation}
\hat{n}=\begin{pmatrix}\sin\theta\cos\phi e^{i\alpha_{bd}}\\
\sin\theta\sin\phi e^{i\alpha_{bs}}\\
\cos\theta
\end{pmatrix},\label{eq:VersorDef}
\end{equation}
where the angles and phases can be chosen to lie in the following range:
\be
	\theta \in \left[0, \frac{\pi}{2} \right]~, \quad
	\phi \in \left[0, 2\pi \right)~, \quad
	\alpha_{bd} \in \left[-\frac{\pi}{2}, \frac{\pi}{2} \right]~, \quad
	\alpha_{bs} \in \left[-\frac{\pi}{2}, \frac{\pi}{2} \right]~. \quad
	\label{eq:Ranges}
\ee
The values of the angles and phases associated to specific directions in flavor space (up and down quarks) are collected in Table~\ref{tab:quarkDir} and shown in the corresponding Figure.
 
\begin{table}[t]
\centering
\renewcommand{\arraystretch}{1.35}
\begin{tabular}{c | c |c c c c}
quark & $\hat n$ & $\phi$ & $\theta$ & $\alpha_{bd}$ & $\alpha_{bs}$ \\\hline
\textcolor{blue}{down} & $(1,0,0)$ & $0$ & $\pi/2$ & $0$ & $0$ \\
\textcolor{ForestGreen}{strange} & $(0,1,0)$ & $\pi/2$ & $\pi/2$ & $0$ & $0$ \\
bottom & $(0,0,1)$ & $0$ & $0$ & $0$ & $0$ \\
\textcolor{orange}{up} & $e^{i \arg(V_{ub})}(V_{ud}^*, V_{us}^*, V_{ub}^*)$ & $0.23$ & $1.57$ & $-1.17$ & $-1.17$ \\
\textcolor{BlueViolet}{charm} & $e^{i \arg(V_{cb})}(V_{cd}^*, V_{cs}^*, V_{cb}^*)$ & $1.80$ & $1.53$ & $-6.2\times 10^{-4}$ & $-3.3\times 10^{-5}$ \\
\textcolor{Cyan}{top} & $e^{i \arg(V_{tb})}(V_{td}^*, V_{ts}^*, V_{tb}^*)$ & $4.92$ & $0.042$ & $-0.018$ & $0.39$
 \end{tabular}\\
 \includegraphics[width=0.5\hsize]{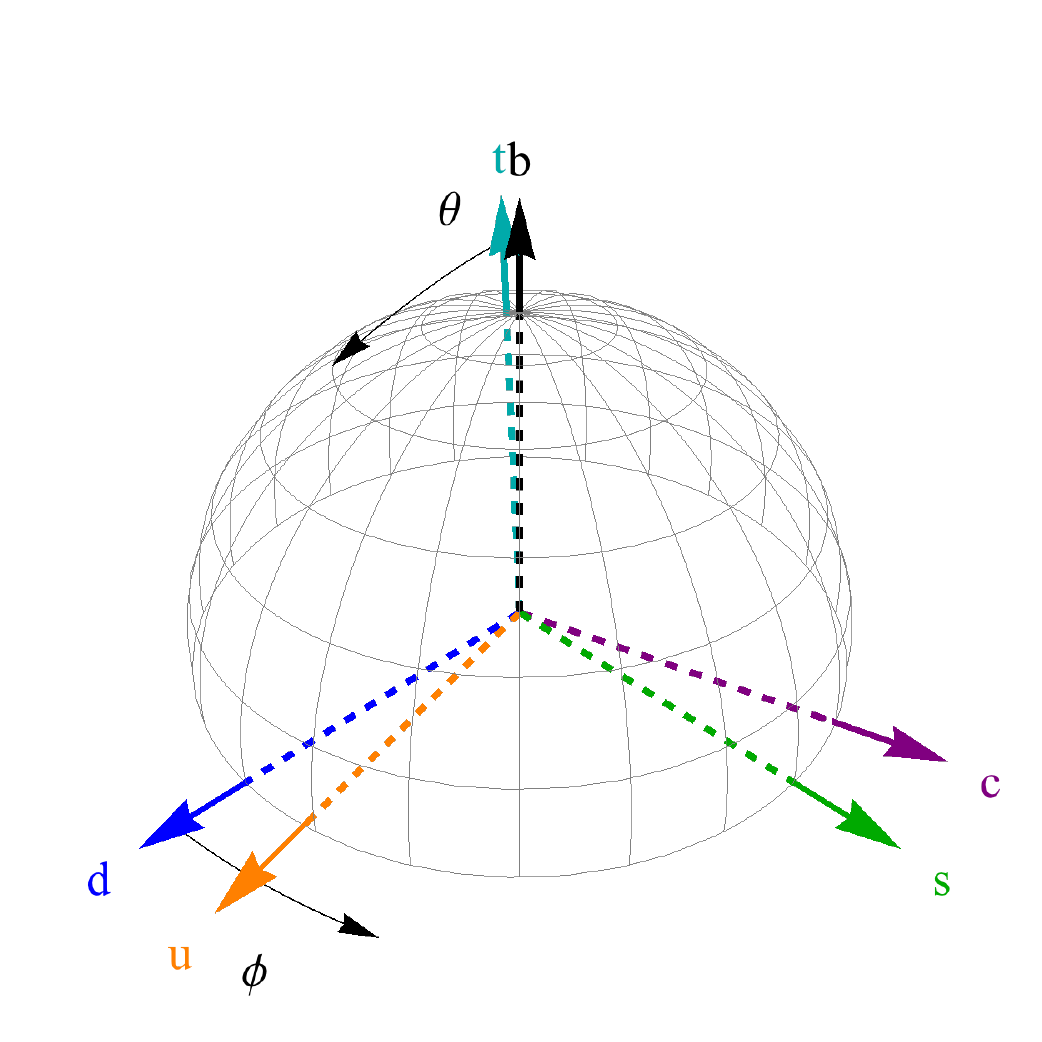}
 \caption{\label{tab:quarkDir} SM quark directions of the unitary vector $\hat{n} _i$. The plot shows the corresponding directions in the semi-sphere described by the two angles $(\theta, \phi)$.}
\end{table}

\begin{table}[t]
\begin{centering}
\begin{tabular}{cc}
\toprule 
Channel & Coefficient dependencies\tabularnewline
\midrule
\midrule 
$d_{i}\to d_{j}\mu^{+}\mu^{-}$ & $C_{S}+C_{T},\ C_{R}$\tabularnewline
\midrule 
$u_{i}\to u_{j}\overline{\nu_{\mu}}\nu_{\mu}$ & $C_{S}+C_{T}$\tabularnewline
\midrule 
$u_{i}\to u_{j}\mu^{+}\mu^{-}$ & $C_{S}-C_{T},\ C_{R}$\tabularnewline
\midrule 
$d_{i}\to d_{j}\overline{\nu_{\mu}}\nu_{\mu}$ & $C_{S}-C_{T}$\tabularnewline
\midrule 
$u_{i}\to d_{j}\mu^{+}\nu_{\mu}$ & $C_{T}$\tabularnewline
\bottomrule
\end{tabular}
\par\end{centering}
\caption{\label{tab:WC dependencies}Dependencies of various semileptonic processes on the three coefficients
$C_{S,T,R}$ (cf. Eq. \eqref{eq:WCmatrix parametrization}). Here and in the text, a given quark level process represents all processes obtained through a crossing symmetry from the shown one.}
\end{table}

The ROFV structure of the semileptonic operators, Eq.~\eqref{eq:WCmatrix parametrization}, implies the existence of correlations between the NP contributions to $b\to s \mu \mu$ anomalous observables and to other observables. In the SMEFT, additional correlations follow from the SU(2)$_L$ invariance of the lagrangian in \Eq{SMEFTlagrangian}. We can then take advantage of the experimental constraints on those additional observables to constrain the flavor directions $\hat n$ accounting for the anomalies. In order to do that, we proceed as follows: for a given direction $\hat {n}$, we fix (some combination of) the overall  coefficients in Eq. \eqref{eq:WCmatrix parametrization} by matching with the
best-fit value of the $C_L^{sb}$ (or $C_9^{sb}$) coefficient obtained from global fits.
Once this is done, we can compute NP contributions to other semileptonic processes as functions of $\hat {n}$, and compare with the corresponding experimental values/bounds. By this procedure, we are able to narrow down considerably the space of allowed flavor directions $\hat {n}$.\footnote{We checked explicitly that the results obtained in this way, i.e by fixing $C^{bs}_{L,9}$ to its best-fit point, or by performing a global $\chi^2$ analysis to get the $95\%$CL excluded region agree very well with each other.}

We analyse the constraints on the direction $\hat n$ under different assumptions. We begin in Sec.~\ref{sec:GeneralCorrelations} by using the effective description in \Eq{NPEFT} and focussing on the case $C_R = 0$. This allows us to derive general correlations with other $d_i d_j \mu \mu$ observables.
In Sec.~\ref{sec:Mediators} we extend the analysis to $\SU(2)_L \times \U(1)_Y$ invariant operators, thus enabling us to consider also observables with up-quarks and/or muon neutrinos. Tab.~\ref{tab:WC dependencies} shows the dependencies of the various types of process upon the three coefficients $C_{S,T,R}$. In particular, we consider specific combinations of $C_{S,T,R}$ obtained in some single-mediator simplified models: $S_3$ and $U_1^\mu$ leptoquarks, as well as of a $Z^\prime$ coupled to the vector-like combination of muon chiralities.
In Sec.~\ref{sec:FlavorSymmetry} we study the connection of our rank-one assumption with $\U(3)^5$ and $\U(2)^5$ flavor symmetries. A discussion on the impact of future measurements is presented in Sec.~\ref{sec:Prospects}, and we conclude in Sec.~\ref{sec:Conclusions}. A simplified fit of the $R_K$ and $R_{K^{*}}$ anomalies as well as some details on the flavor observables considered in this work, are collected in two Appendices.

\section{General correlations in V-A solutions}
\label{sec:GeneralCorrelations}

In this Section, we study the correlations that follow directly from the rank-one condition, for all models in which NP couples only to left-handed fermions. We begin by using the effective description in \Eq{NPEFT}. 
For $C_R=0$, and for fixed $\theta$ and $\phi$ in the ranges specified by Eq. \eqref{eq:Ranges}, the coefficient  $C_L = C_S + C_T$ and the phase $\alpha _{bs}$ are univocally determined by the $b\to s\mu^+\mu^-$ anomalies fit:
\be 
C_L \sin \theta \cos \theta \sin \phi e^{i\,\alpha_{bs}}= C_L ^{bs} \equiv \frac{e^{i \alpha_{bs}}}{\Lambda_{bs}^2}~.
\label{Eq:CLmatchingfit}
\ee
From a fit of the observables listed in Table~\ref{tab:bsmumuObs}, described in detail in  App.~\ref{app:bsmumu}, we find that the phase $\alpha_{bs}$ has an approximately flat direction in the range $|\alpha_{bs}| \lesssim \pi/4$. Since a non-zero phase necessarily implies a lower $\Lambda_{bs}$ scale in order to fit the anomalies, to be conservative we fix $\alpha_{bs} = 0$. In this case the best-fit point for the NP scale is
\be
(\Lambda_{bs})^{\text {best-fit}}=38.5\,\TeV \qquad (\alpha_{bs}\equiv 0). 
\label{Eq:CLfitArgZero}
\ee

\begin{table}[t]
\begin{centering}
\begin{tabular}{|c|c|c|c|}
\hline 
Observable & Experimental value/bound & $\text{SM}$ prediction & References\tabularnewline
\hline 
\hline 
$\text{Br}(B_{d}^0 \to\mu^{+}\mu^{-})$ & $<2.1\times10^{-10}\ \text{(95\% CL)}$ & $(1.06\pm0.09)\times10^{-10}$ & \cite{Aaboud:2018mst,Bobeth:2013uxa}\tabularnewline
\hline 
$\text{Br}(B^{+}\to\pi^{+}\mu^{+}\mu^{-})_{\left[1,6\right]}$ & $(4.55_{-1.00}^{+1.05}\pm0.15)\times10^{-9}$ & $(6.55\pm1.25)\times10^{-9}$ & \cite{Aaij:2015nea,Du:2015tda,Khodjamirian:2017fxg}\tabularnewline
\hline 
$\text{Br}(K_{S}\to\mu^{+}\mu^{-})$ & $<2.4\times10^{-10}\ \text{(95\% CL)}$ & $ (5.0\pm 1.5)\times 10^{-12}$ & \cite{Aaij:2017tia,LHCb:2019aoh}\tabularnewline
\hline 
$\text{Br}(K_{L}\to\mu^{+}\mu^{-})_{\text{SD}}$ & $<2.5\times10^{-9}\ $ & $\approx0.9\times10^{-9}$ & \cite{Ambrose:2000gj,Isidori:2003ts}\tabularnewline
\hline 
$\text{Br}(K_{L}\to \pi^0\mu^{+}\mu^{-})$ & $<3.8\times10^{-10}\ \text{(90\% CL)}$ & $1.41^{+0.28}_{-0.26} (0.95^{+0.22}_{-0.21}) \times 10^{-11}$ & \cite{AlaviHarati:2000hs,DAmbrosio:1998gur,Buchalla:2003sj,Isidori:2004rb,Mescia:2006jd}\tabularnewline
\hline
\end{tabular}
\par\end{centering}
\caption{\label{tab:Correlated-observables}Observables with direct correlation with $bs\mu\mu$.}
\end{table}

\begin{figure}[t]
\centering
\includegraphics[width=0.49\hsize]{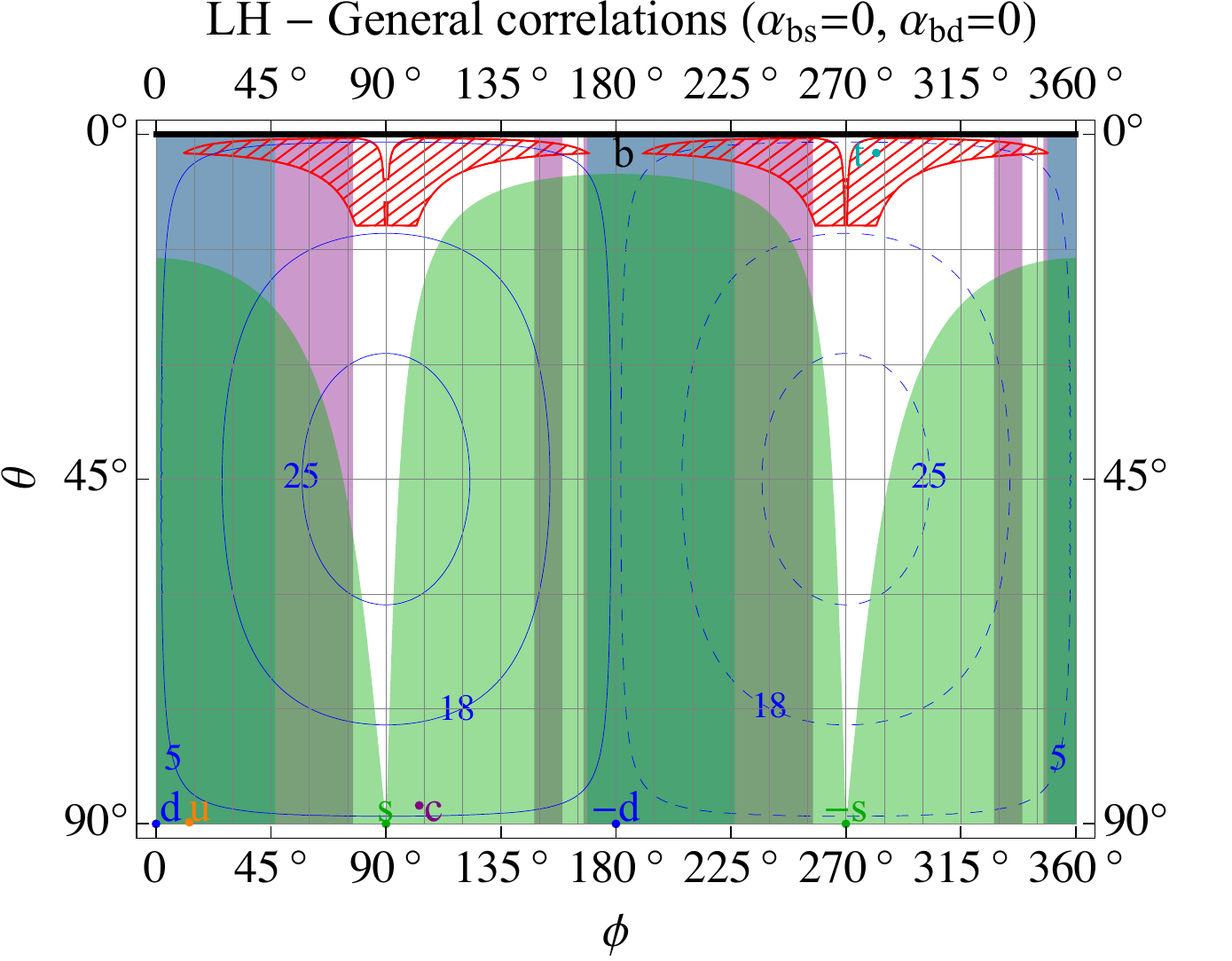}
\includegraphics[width=0.49\hsize]{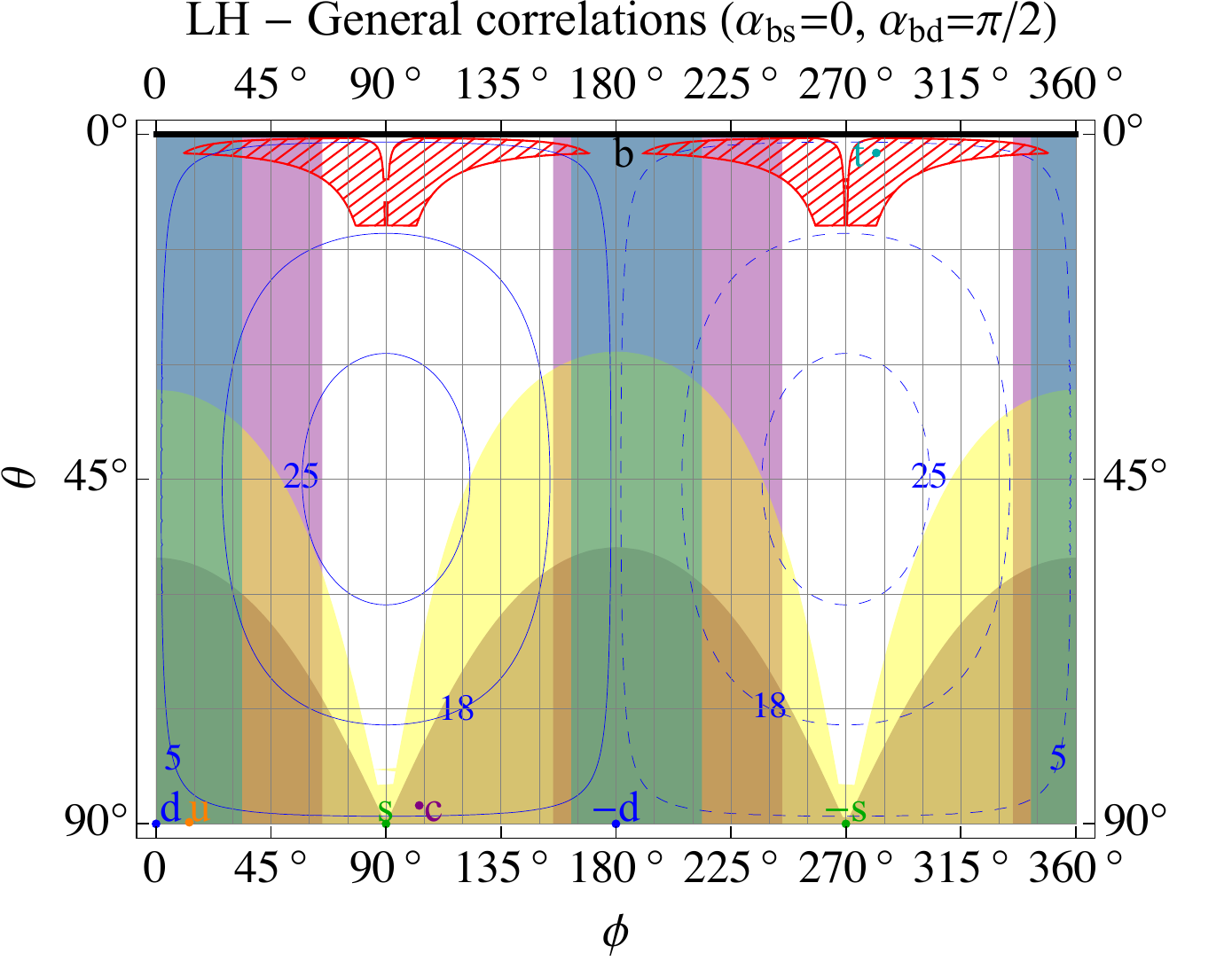}\\
\includegraphics[width=0.5\hsize]{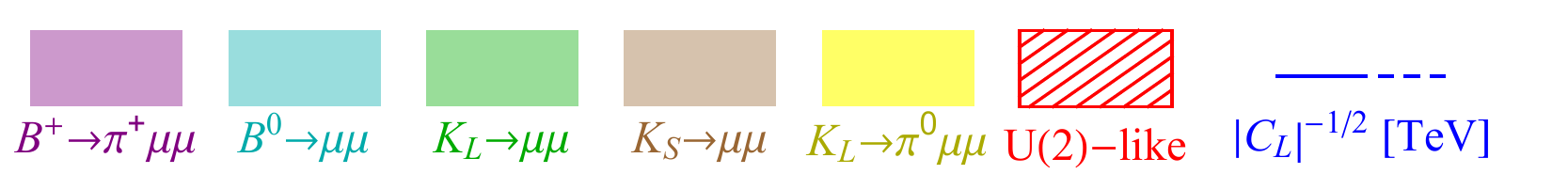}
\caption{Limits in the plane $(\phi, \theta)$ for two choices of the phases $\alpha_{bs}$ and $\alpha_{bd}$ from observables with direct correlation with $R_{K^{(*)}}$. The blue contours correspond to the value of $|C_L|^{-1/2}$ in TeV, where solid (dashed) lines are for positive (negative) $C_L$. The meshed red region correspond to the one suggested by partial compositeness or $SU(2)_q$-like flavor symmetry, Eq.~\eqref{eq:TopLikeDirection} with $|a_{bd,bs}| \in [0.2 - 5]$.}\label{fig:Minimal}
\end{figure}

We now constrain $\hat n$ (or, more precisely, $\theta$ and $\phi$ for given $\alpha_{bs}$) using the other observables correlated with $R_K$ by the relation $C^{ij}_L = C_L \hat n_i \hat n^*_j$. Such observables are associated to the  quark-level transitions 
\begin{equation}
d_{i}\to d_{j}\mu^{+}\mu^{-}  
\label{eq:minimalconstraints}
\end{equation}
and cross-symmetric counterparts. The most relevant among those observables are listed in Tab.~\ref{tab:Correlated-observables}, and the corresponding allowed regions for $\theta$, $\phi$ are shown in Fig.~\ref{fig:Minimal} in the two cases $(\alpha_{bd},\alpha_{bs})=(0,0)$ and $(\alpha_{bd},\alpha_{bs})=(\pi/2,0)$. 

As can be seen from the plots, the most severe bounds arise from $B^+ \to \pi ^+ \mu ^+ \mu ^-$ (LHCb \cite{Aaij:2015nea}) and $K_L \to \mu ^+ \mu ^-$  (E871 \cite{Ambrose:2000gj,Isidori:2003ts}). However, the latter observable does not yield any bound for $\alpha _{bd}-\alpha _{bs}=\pi/2$, i.e. for $\text{Re}\, C_L ^{ds}=0$. The imaginary part of that coefficient can instead be tested by $K_S \to \mu^+ \mu^-$ (LHCb \cite{Aaij:2017tia,LHCb:2019aoh,Aaij:2012rt}) and $K_L \to \pi^0 \mu^+ \mu^-$ (KTeV \cite{AlaviHarati:2000hs}).\footnote{We thank C. Bobeth for suggesting us to consider this observable.}
More details on the observables and their NP dependence can be found in App.~\ref{App:Flavor observables}.

As a final remark, let us stress that here and in the following we are ignoring possible NP contributions to (pseudo)scalar, tensor, or dipole operators. While these are known to be too constrained to
give significant contributions to $R_{K^{(*)}}$ (see e.g. \cite{Hiller:2014yaa}),
they may nonetheless produce important effects in other observables,
so that some of the bounds discussed here may be relaxed, if some
degree of fine-tuning is allowed.

\section{SMEFT and Simplified mediators}
\label{sec:Mediators}

Let us now assume that the effective operators in \Eq{NPEFT} originate from the SM-invariant ones in \Eq{SMEFTlagrangian}, as expected. The SU(2)$_L$ invariance then relates the processes $d_{i}\to d_{j}\mu^{+}\mu^{-}$ in \Eq{minimalconstraints} to the processes involving up quarks and muon neutrinos listed in \Tab{WC dependencies}.
Using the experimental constraints on those, we can impose further constraints on $\hat n$.
These, though, are model dependent even in the $C_R = 0$ case, as they depend on the relative size of the two operators in \Eq{SMEFTlagrangian} contributing to $C_L$, i.e. $C_S$ and $C_T$. 
The origin of the model dependence can be clarified taking advantage of a phenomenological observation. Our analysis (see below) shows that the most relevant constraints come from the processes 
\begin{equation}
d_{i}\to d_{j}\mu^{+}\mu^{-}  \quad \text{and} \quad d_{i}\to d_{j}\overline{\nu_{\mu}}\nu_{\mu} \;.
\label{eq:generalconstraints}
\end{equation}
As Table \ref{tab:WC dependencies} shows, those two classes of processes are associated respectively to
the two operators $\mathcal{O}^\pm$ whose Wilson coefficients are $C_\pm = C_S\pm C_T$ (note that $C_+ \equiv C_L$), i.e.
\begin{align}
\mathcal{O}^+_{ij} &= \frac{\mathcal{O}^S_{ij}+\mathcal{O}^T_{ij}}{2} = \left(  \overline{q_{iL}} \gamma_\mu \ell_{2L} \right)\left( \overline{\ell_{2L}} \gamma^\mu  q_L^j\right)~, \\
\mathcal{O}^-_{ij} &= \frac{\mathcal{O}^S_{ij}-\mathcal{O}^T_{ij}}{2} = 2 \left( \overline{q^{c}_{iL}} \ell_{2L} \right) \left( \overline{\ell_{2L}}  q_{jL}^c \right) \;.
\label{eq:Opm}
\end{align}
The model-independent constraints shown in \Fig{Minimal} only take into account the $d_{i}\to d_{j}\mu^{+}\mu^{-}$ processes and as such only depend on $C_+$, which is thus the only combination fixed by $R_{K^{(*)}}$. On the other hand, the model-dependent weight of the $d_{i}\to d_{j}\overline{\nu_{\mu}}\nu_{\mu}$ constraints depends on the relative size of $C_-$. In this context, the results in \Fig{Minimal} correspond to a SMEFT with $C_- = 0$, i.e.\ to the $U_1$ case in \Tab{Mediators}.

Note that the experimental constraints on the processes involving neutrinos do not distinguish among the three neutrino flavors. In order to get constraints on the muon neutrino operators we consider, one should then make an assumption on the relative size of the operators with different neutrino flavors. Below, we will conservatively assume that only the muon neutrino operators contribute to the neutrino processes.

In order to reduce the number of free parameters, we focus in this Section on single-mediator simplified models, which generate specific combinations of the three operators when integrated out at the tree-level. Some relevant benchmarks are shown in Table \ref{tab:Mediators}, where in the last column we list the ratios:
\begin{equation}
c_X\equiv \dfrac{C_X}{C_S+C_T} = \frac{C_X}{C_+} \qquad (X=S,T,R)~.
\label{eq:c_x ratios}
\end{equation}

Notice that the exclusions shown in Fig.~\ref{fig:Minimal} hold in all models in Table \ref{tab:Mediators}, except for the $Z_{V}'$ which has vector-like coupling to muons.
We find that the most relevant bounds, beyond those already analized, arise from the FCNC observables $\text{Br}(K^+\to \pi ^+\nu_{\mu}\overline{\nu_{\mu}})$ and $\text{Br}(K_L \to \pi ^0\nu_{\mu}\overline{\nu_{\mu}})$, reported in \Tab{Correlated-observables-simplified}. The connection of these observables with the B-meson anomalies has also been emphasised in Ref.~\cite{Bordone:2017lsy}. The effect of constraints from rare kaon decays on LQ models addressing instead the $\epsilon^\prime/\epsilon$ anomaly have been studied in Ref.~\cite{Bobeth:2017ecx}. Some simplified models also allow to compute neutral meson mixing amplitudes, which we include in the analysis when appropriate.

\begin{table}[t]
\centering
\begin{equation*}
\begin{array}{c|c|c|c}
\textrm{Simplified model} & \textrm{Spin} & \textrm{SM irrep} & (c_S,c_T, c_R)  \\
\hline
S_3 & 0 & (\overline{3},3,1/3) &  (3/4,1/4,0) \\
U_1 & 1 & (3,1,2/3) & (1/2,1/2,0) \\
U_3 & 1 & (3,3,2/3) &(3/2,-1/2,0) \\
V' & 1 & (1,3,0) & (0,1,0) \\
Z' & 1 & (1,1,0) &(1,0,c_R) 
\end{array}
\end{equation*}\vspace{-0.5cm}
\caption{\label{tab:Mediators} Wilson coefficients ratios (cf. Eq. \eqref{eq:c_x ratios}) for some single-mediator simplified models.}
\end{table}

\begin{table}[t]
\centering
\begin{tabular}{|c|c|c|c|}
\hline 
Observable & Experimental value/bound & $\text{SM}$ prediction & References\tabularnewline
\hline 
\hline 
$\text{Br}(K^+\to \pi ^+\nu_{\mu}\overline{\nu_{\mu}})$ & $< 2.44 \times 10^{-10}\ \text{(95\% CL)}$ & $(8.4\pm 1.0)\times 10^{-11}$ & \cite{NA62:2019ajt,Buras:2015qea}\tabularnewline
\hline
$\text{Br}(K_L \to \pi ^0\nu_{\mu}\overline{\nu_{\mu}})$ & $<3.0\times10^{-9}\ \text{(90\% CL)}$ & $ (3.4\pm 0.6)\times 10^{-11}$ & \cite{Ahn:2018mvc,Buras:2015qea}\tabularnewline 
\hline
\end{tabular}
\caption{\label{tab:Correlated-observables-simplified}$R_{K^{(*)}}$-correlated observables for single-mediator models.}
\end{table}

Some comments are in order regarding the phenomenological relevance of the various processes listed in \Tab{WC dependencies}. Flavor observables of the type $u_i\to u_j \overline {\nu _\mu} \nu _\mu $ or $u_i\to u_j \mu ^+ \mu ^- $ are much less constrained than their $d_i \to d_j$ counterparts, from the experimental point of view. On the other hand, the charged current processes $u_{i}\to d_{j}\mu^{+}\nu_{\mu}$ (which could in principle yield correlations between $C_+$ and $C_-$), being unsuppressed in the SM, receive only tiny corrections in the present framework. It turns out that all these observables lead to weaker bounds than those arising from other sectors, so that we omit them altogether from our analysis (as possibly relevant observables, we examined $\text{Br}(J/\psi \to \text{invisible})$, $\text{Br}(D^0 \to \mu ^+ \mu ^-)$ and $\text{Br}(K^+(\pi ^+) \to \mu ^+ \nu _\mu)$, for the three kind of quark level processes mentioned above, respectively.). Instead, for the purpose of comparison, we display in this Section the collider bounds arising from the high-$p_T$ tails of muonic Drell-Yan process measured at LHC \cite{Aaboud:2017buh}, for which we follow the analysis of Ref. \cite{Greljo:2017vvb}. As it can be seen from the plots below, the collider bounds are outmatched by FCNC bounds in a large part of parameter space. The only region where LHC searches are the most relevant constraint is close to the bottom quark direction, i.e. for $\theta \ll 1$, as it can be seen directly in the top-left panel of Fig.~\ref{fig:S3limits} in the case of the $S_3$ leptoquark.

In the rest of this Section we focus on the following models: 

\begin{enumerate}

\item \textbf{Scalar leptoquark $S_3$.} This is the simplest \textit{renormalizable} model explaining the $R_{K^{(*)}}$ anomalies with NP in muons \cite{Gripaios:2014tna,Hiller:2014yaa,Varzielas:2015iva,Hiller:2017bzc,Dorsner:2017ufx,Cline:2017aed,Cline:2017qqu}.
\item \textbf{Vector leptoquark $U_1$.} Besides having some theoretical motivation (from Pati-Salam SM extensions), this is the \textit{only} single-mediator simplified model for which a combined explanation of $R_{K^{(*)}}$ and $R_{D^{(*)}}$ anomalies is possible \cite{Barbieri:2015yvd,Alonso:2015sja,Barbieri:2016las,Buttazzo:2017ixm,DiLuzio:2017vat,Barbieri:2017tuq,Bordone:2017bld,DiLuzio:2018zxy,Bordone:2018nbg,Crivellin:2018yvo,Baker:2019sli}.
\item \textbf{Vector singlet $Z'$} with vector-like coupling to muons and a single vectorlike partner for quark doublets. Arguably the most compelling $\mathcal{O}_9$-type solution, it is relevant to some interesting proposals such as gauged $L_\mu - L_\tau$ or $B_3 - 3 L_\mu$, see for example Refs.~\cite{Altmannshofer:2014cfa,Crivellin:2015mga,Bonilla:2017lsq,Biswas:2019twf}.

\end{enumerate}

\subsection{Scalar leptoquark $S_3$}

\begin{figure}[t]
\centering
\hspace{-0.5cm}\includegraphics[width=0.46\hsize]{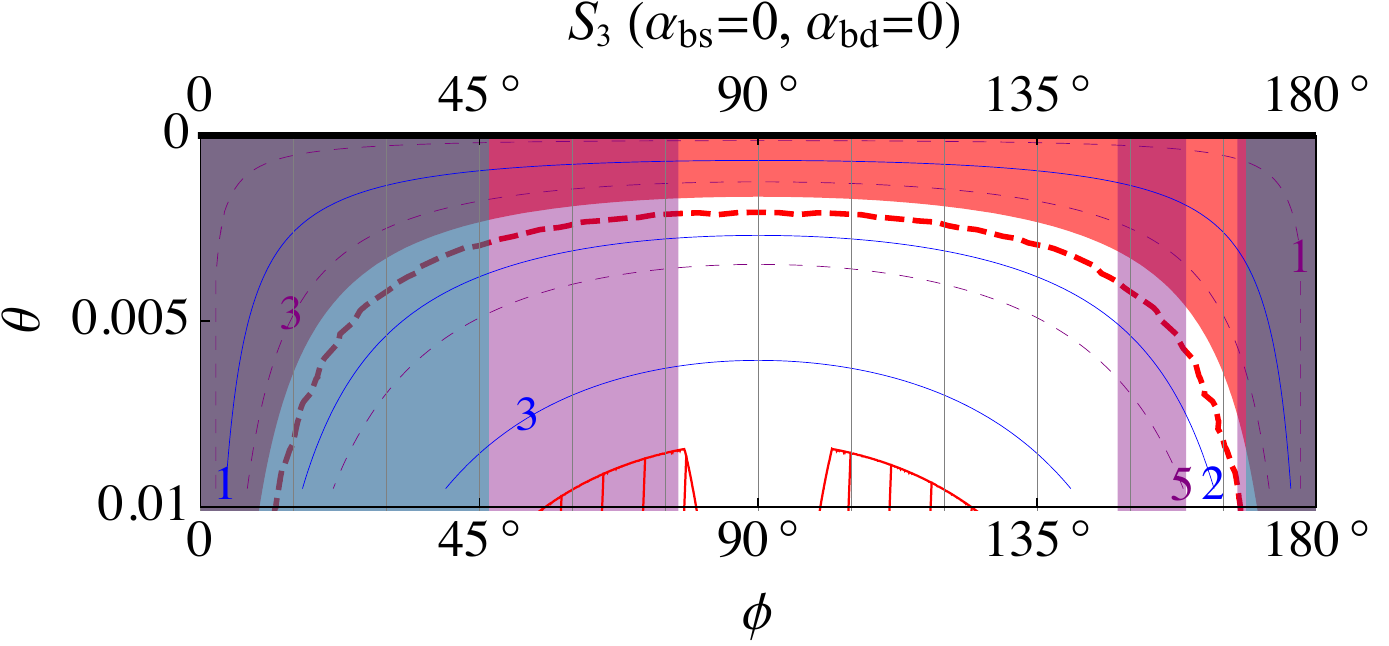} \quad
\includegraphics[width=0.48\hsize]{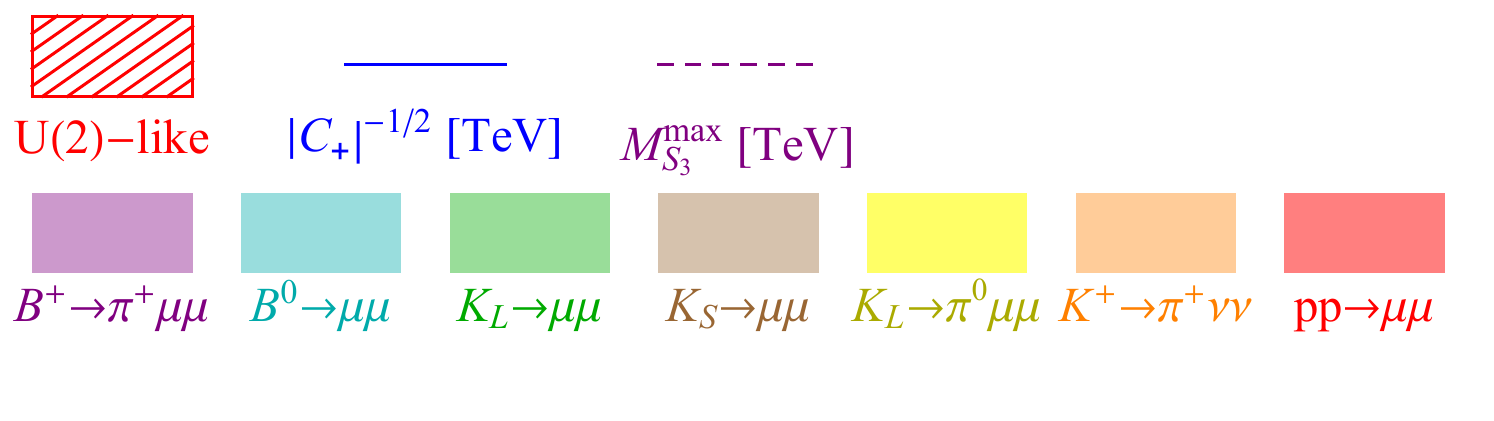}\\
\includegraphics[width=0.48\hsize]{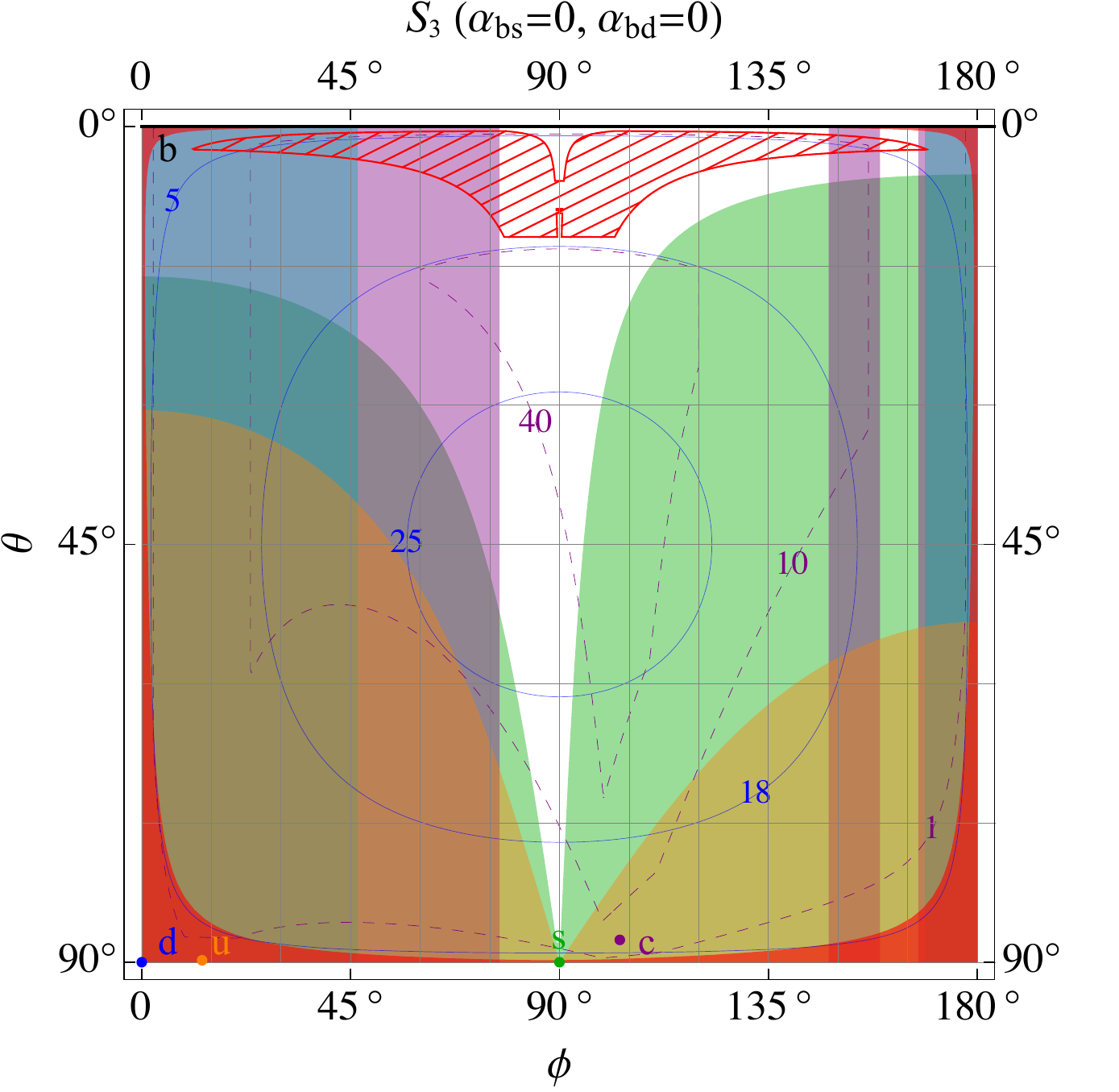}
\includegraphics[width=0.48\hsize]{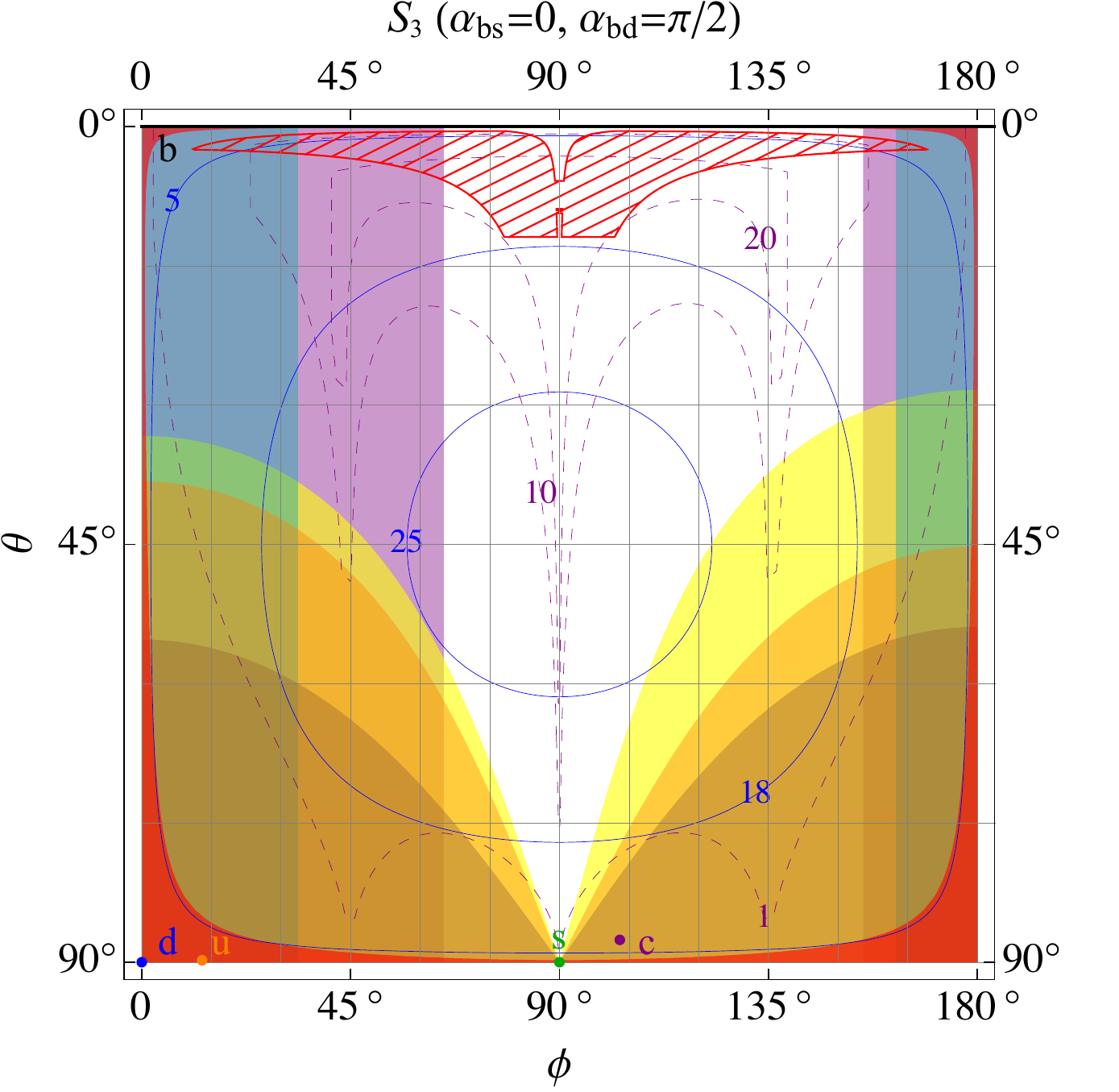}
\caption{Limits in the plane $(\phi, \theta)$ for the scalar leptoquark $S_3$ and two choices of the phases $\alpha_{bs}$ and $\alpha_{bd}$. In addition to the limits in Fig.1, the orange bound is from $K \to \pi \nu\nu$ while the red one is from the high-$p_T$ tail of $p p \to \mu^+ \mu^-$ at the LHC \cite{Greljo:2017vvb}. The top-left panel is a zoom of the region $\theta \ll 1 0$ of the bottom-left one, which shows in more detail the region excluded by LHC dimuon searches. The dashed purple contour lines are the upper limits (in TeV) on the leptoquark mass from $\Delta F = 2$ processes.}\label{fig:S3limits}
\end{figure}

The relevant interaction of the $S_3$ leptoquark with SM quarks and leptons can be described by the Lagrangian 
\be
	\LL_{\rm NP} \supset \beta_{i \mu} (\overline{q^{c}_{iL}} \epsilon \sigma^a \ell_{2L}) \, S_3^a + {\rm h.c.} ~,
	\label{eq:S3Lagr}
\ee
where we focussed on the interaction with muons. It is clear that this model falls under the category described by Eq.~\eqref{eq:linearNPlagrangian}.
Integrating out $S_3$ at the tree-level, the effective operators in Eq.~\eqref{eq:SMEFTlagrangian} are generated, with
\be
	C_S^{ij} = \frac{3}{4} \frac{\beta_{i \mu}^* \beta_{j \mu}}{M_{S_3}^2}~, \quad
	C_T^{ij} = \frac{1}{4} \frac{\beta_{i \mu}^* \beta_{j \mu}}{M_{S_3}^2}~, \quad
	C_R^{ij} = 0~.
	\label{eq:S3SMEFTmatch}
\ee
We can match to our parametrization by writing the coupling in \Eq{S3Lagr} as $\beta_{i \mu}^* \equiv \beta^* \, \hat n_i$, giving $C_+ = |\beta|^2 / M_{S_3}^2 > 0$, $c_S = 3/4$, $c_T = 1/4$, and $c_R = 0$.
Since in this case $C_+ = C_L > 0$ and the r.h.s. of Eq.~\eqref{Eq:CLmatchingfit} is also positive, the angle $\phi$ is restricted to the range $\phi \in [ 0, \pi ]$.
The constraints on $\phi$ and $\theta$ we obtain are shown in \Fig{S3limits}.

The scalar LQ $S_3$ generates a contribution to $\Delta F = 2$ processes at the one-loop level. The relevant diagrams are finite and the contribution from muonic loops is given by
\be
	\Delta \LL_{\Delta F = 2} =  - \frac{5}{128 \pi^2} C_+^2 M_{S_3}^2 \left[ (\hat n_i \hat n_j^* \overline{d_{iL}} \gamma^\alpha d_{jL})^2 + ( V_{i k} \hat n_k \hat n_l^* V^*_{j l} \; \overline{u_{iL}} \gamma^\alpha u_{jL})^2 \right]~.
	\label{eq:DeltaF2_S3}
\ee
Given a direction in quark space, i.e.\ a fixed $\hat n$, and fixing $C_+$ to reproduce $R_{K^{(*)}}$, the experimental bounds on $K - \bar K$, $B_{d,s} - \bar B_{d,s}$, and $D_0 - \bar D_0$ mixing can be used to set an \emph{upper limit} on the LQ mass, assuming the muonic contributions shown in Eq.~\eqref{eq:DeltaF2_S3} to be dominant compared to other possible NP terms. For the sake of clarity, it is worth remarking that loops involving $\tau$ leptons could in general also give substantial contributions to Eq.~\eqref{eq:DeltaF2_S3}, possibly making the bounds on $M_{S_3}$ qualitatively stronger or weaker than those shown in Fig.~\ref{fig:S3limits}, depending on the specific flavor structure of leptoquark couplings.
Another upper limit on its mass, for a given value of $C_+$, can be set by requiring that the coupling $\beta$ does not exceed the perturbative unitarity limit $|\beta^{\rm max}|^2 = (8\pi)/(3\sqrt{3})$ \cite{DiLuzio:2017chi}.
The contours of the strongest of these two upper limits on $M_{S_3}$ are shown as dashed purple lines (in TeV) in the plots of Fig.~\ref{fig:S3limits}. The perturbativity limit is never stronger than the one from $\Delta F=2$ processes in this scenario. More details are reported in App.~\ref{sec:DeltaF2limits}. Direct searches at the LHC of pair-produced leptoquarks, on the other hand, set \emph{lower limits} on its mass, which are now in the $\sim 1 \TeV$ range.

\subsection{Vector leptoquark $U_1$}
\label{sec:U1}

\begin{figure}[t]
\centering
\includegraphics[width=0.48\hsize]{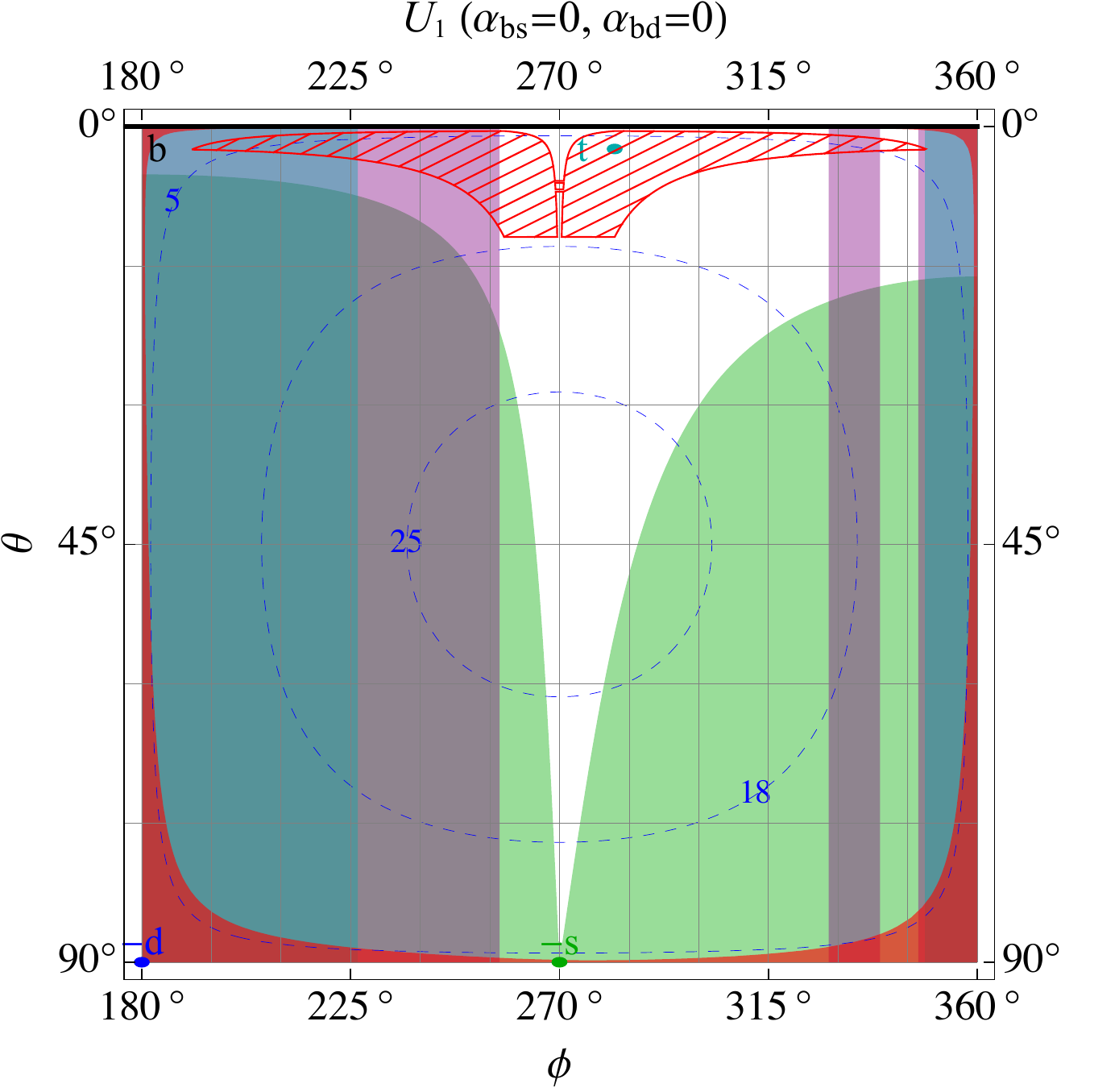}
\includegraphics[width=0.48\hsize]{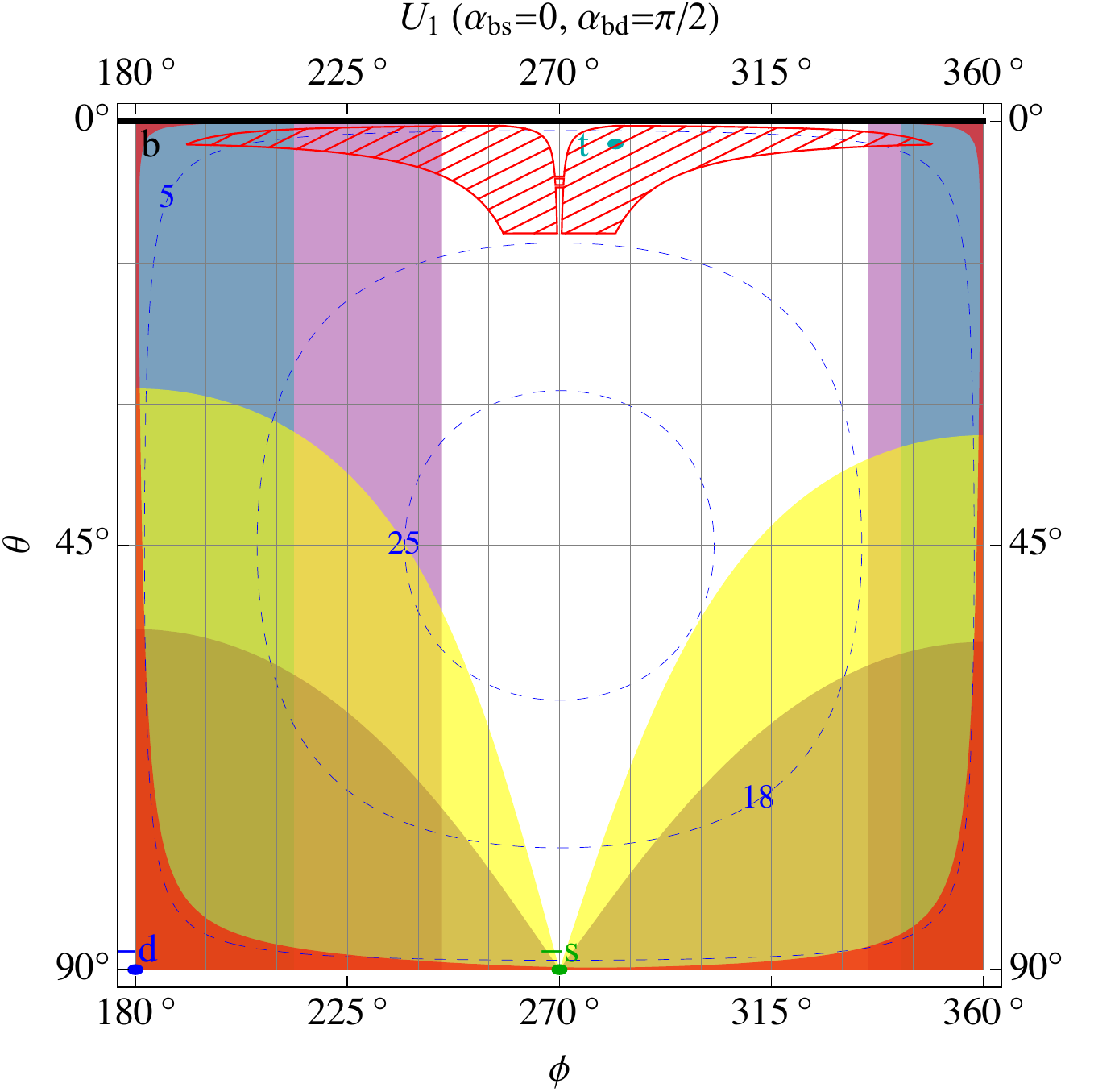}\\
\includegraphics[width=0.63\hsize]{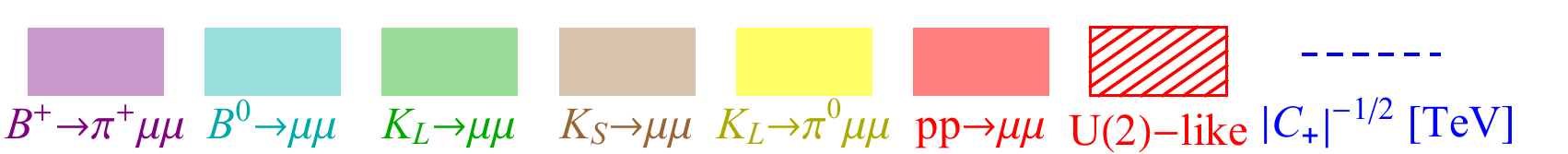}
\caption{Limits in the plane $(\phi, \theta)$ for the vector leptoquark $U_1$ and two choices of the phases $\alpha_{bs}$ and $\alpha_{bd}$. The red region is excluded by the high-$p_T$ tail of $p p \to \mu^+ \mu^-$ at the LHC \cite{Greljo:2017vvb}.}\label{fig:U1limits}
\end{figure}

The interaction lagrangian of the vector leptoquark $U_1$ is
\be
	\LL_{\rm NP} \supset \gamma_{i \mu} (\overline{q_{iL}} \gamma_\alpha \ell_{2  L}) \, U_1^\alpha + {\rm h.c.} ~,
	\label{eq:U1Lagr}
\ee
The matching to the SMEFT operators generated by integrating out $U_1$ at the tree-level is given by
\be
	C_S^{ij} = - \frac{1}{2} \frac{\gamma^{\phantom{*}}_{i \mu} \gamma_{j \mu}^*}{M_{U_1}^2}~, \quad
	C_T^{ij} = - \frac{1}{2} \frac{\gamma^{\phantom{*}}_{i \mu} \gamma_{j \mu}^*}{M_{U_1}^2}~, \quad
	C_R^{ij} = 0~.
	\label{eq:U1SMEFTmatch}
\ee
We can match to our parametrization by defining $\gamma_{i \mu} \equiv \gamma \, \hat n_i$, corresponding to $C_+ = - |\gamma|^2 / M_{U_1}^2 < 0$, $c_S = 1/2$, $c_T = 1/2$, and $c_R = 0$.
Contrary to the $S_3$ model, the $U_1$ LQ implies $C_+ = C_L < 0$. Therefore, Eq.~\eqref{Eq:CLmatchingfit}, whose r.h.s. is positive, restricts the angle $\phi$ to the range $[\pi, 2\pi)$.
The constraints on $\phi$ and $\theta$ we obtain are shown in \Fig{U1limits}. As anticipated, they coincide with the constraints (in the $\pi < \phi < 2\pi$ part) of \Fig{Minimal}.

Like $S_3$, also the $U_1$ vector LQ contributes to meson anti-meson mixing at one-loop. Such a contribution is however UV-divergent and, in order to be calculable, requires a UV-completion of the simplified model. In general such UV completions contains other contributions to the same processes, which must also be taken into account \cite{Barbieri:2016las,DiLuzio:2017vat,Barbieri:2017tuq,Bordone:2017bld,DiLuzio:2018zxy,Bordone:2018nbg,Baker:2019sli}.

\subsection{Vector singlet $Z^\prime$ with vector-like couplings to muons}
\label{sec:Zprime}

\begin{figure}[t]
\centering
\includegraphics[width=0.49\hsize]{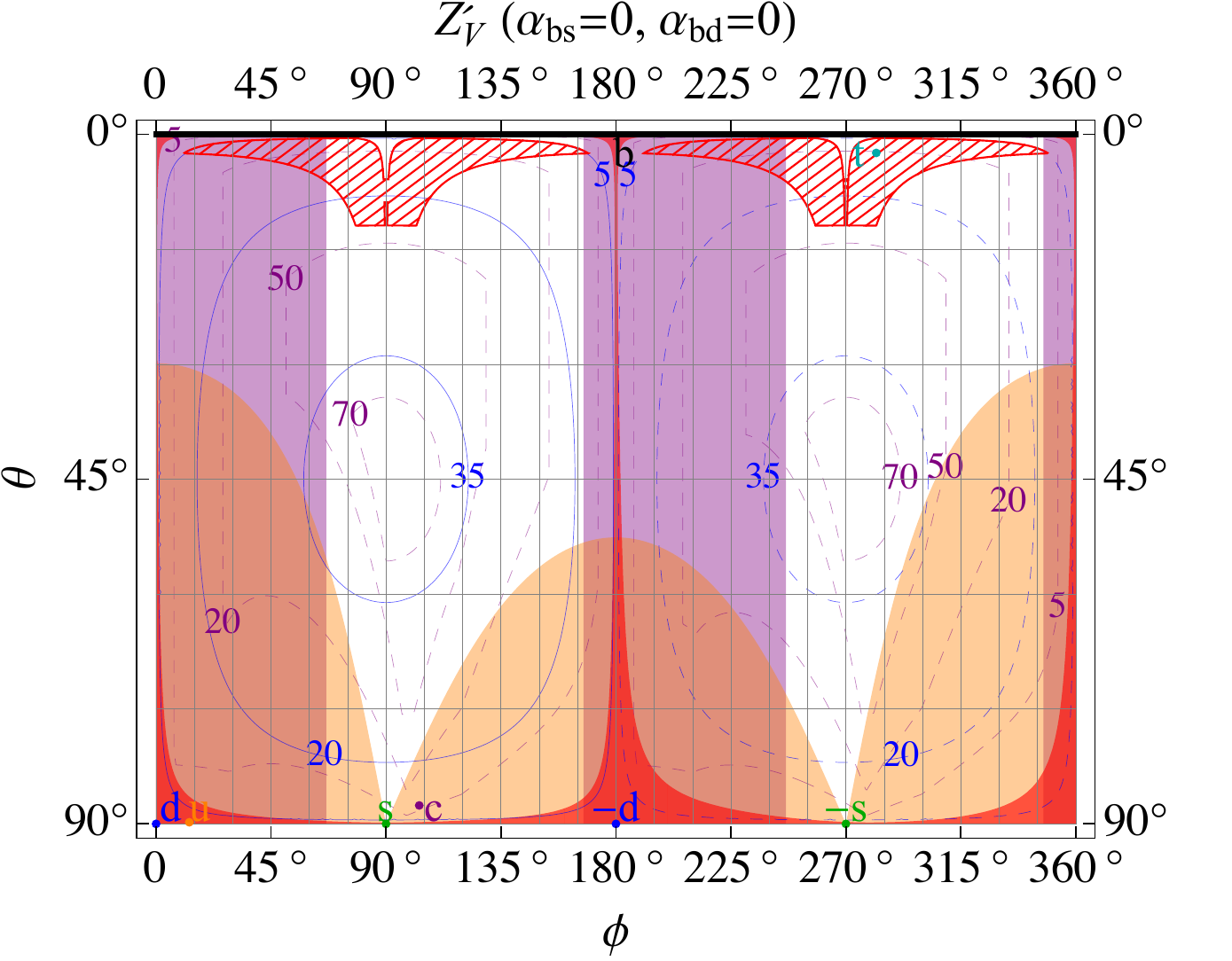}\!\!\!
\includegraphics[width=0.49\hsize]{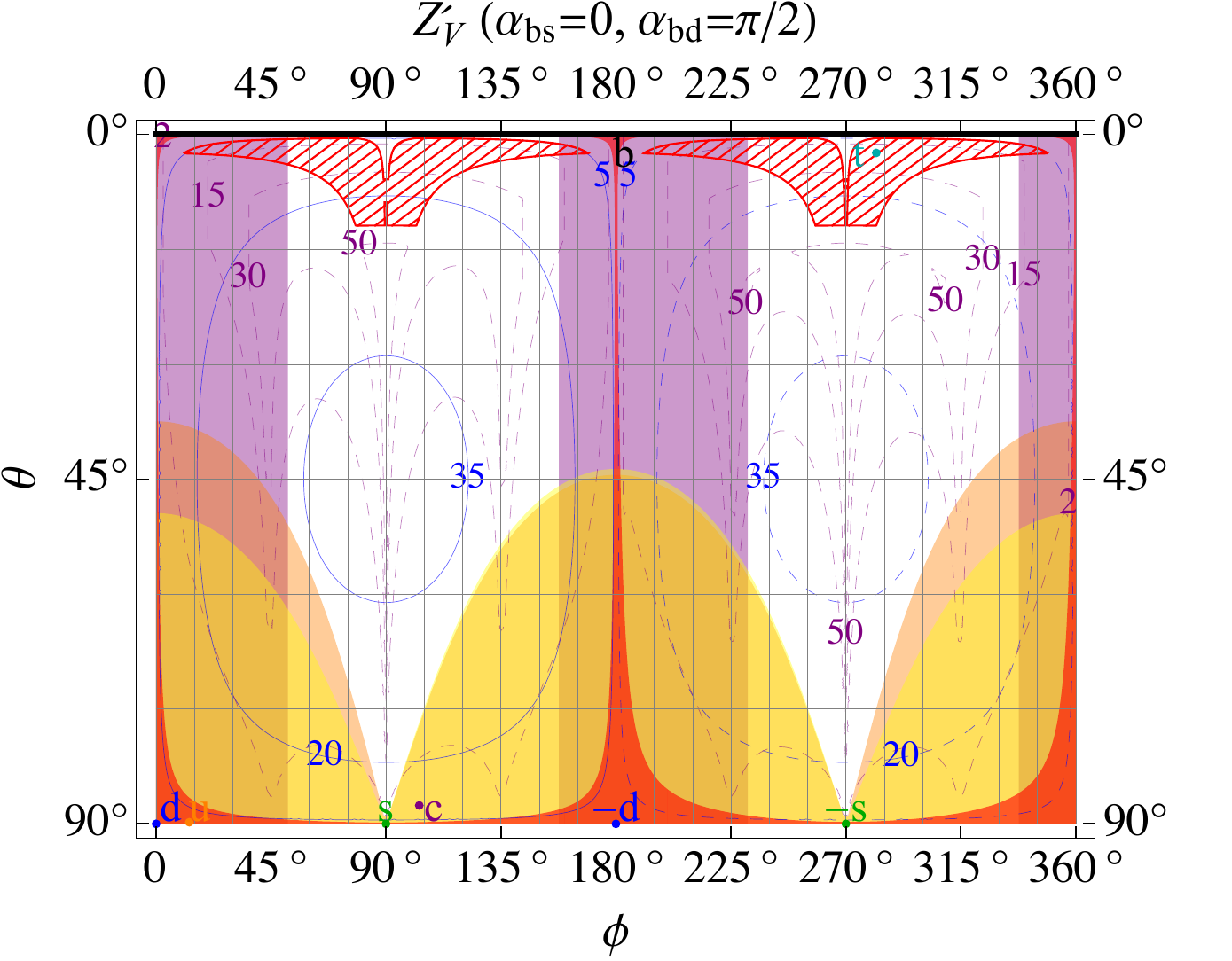}\\
\includegraphics[width=0.53\hsize]{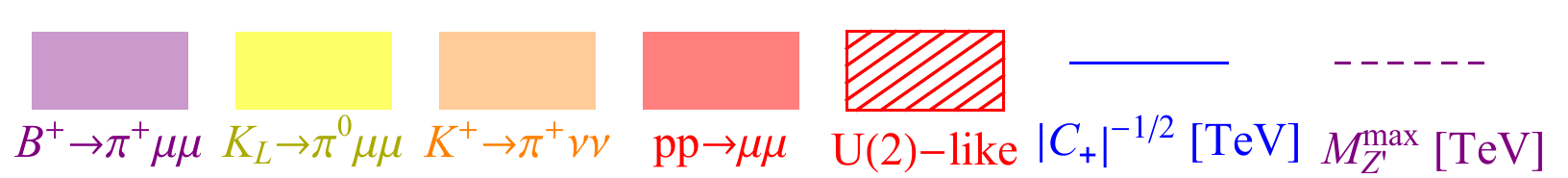}
\caption{Limits in the plane $(\phi, \theta)$ for the vector singlet $Z^\prime$ with vector-like couplings to muons and two choices of the phases $\alpha_{bs}$ and $\alpha_{bd}$. The dashed purple contour lines are upper limits on the $Z^\prime$ mass [TeV] from $\Delta F = 2$ processes using Eq.~\eqref{eq:MaxZpMass}.}\label{fig:Zplimits}
\end{figure}

Let us consider a heavy singlet vector $Z^\prime$ with couplings:
\be
	\LL_{\rm NP} \supset \left[ g^{ij}_q (\overline{q_{iL}} \gamma^\alpha q_{jL}) + g_\mu (\overline{\ell_{2L}}\gamma^\alpha \ell_{2L} + \overline{\mu_R} \gamma^\alpha \mu_R) \right] \, Z^\prime_\alpha \;,
	\label{eq:ZpLagr}
\ee
where $g^{ij}_q = g_q   \hat n_i \hat n_j^*$. Such a flavor structure of the $Z^\prime$ couplings to quarks could arise, for example, by assuming that they couple to $Z^\prime$ only via the mixing with a heavy vector-like quark doublet $Q$ in the form 
\begin{equation}
M_i \overline Q \, q_{iL} + \text{h.c.} \;.
\label{eq:mixing}
\end{equation}
In such a case, $\hat n_i \propto M^*_i$. The matching with the SMEFT operators in this case is given by 
\be
	C_S^{ij} = -\frac{g_q g_\mu}{M_{Z^\prime}^2} \hat n_i \hat n_j^*~, \quad
	C_T^{ij} = 0~, \quad
	C_R^{ij} = -\frac{g_q g_\mu}{M_{Z^\prime}^2} \hat n_i \hat n_j^*~,
	\label{eq:ZpSMEFTmatch}
\ee
corresponding to $C_+ = -g_q g_\mu / (M_{Z^\prime}^2)$, $c_S = 1$, $c_R = 1$, and $c_T = 0$.
The matching to the operators relevant for the $bs\mu\mu$ anomalies is now
\be
	C_+ \sin \theta \cos\theta \sin\phi e^{i \alpha_{bs}} = \frac{G_F \alpha}{\sqrt{2} \pi} V_{tb} V_{ts}^* \Delta C_9^\mu~.
	\label{eq:C9matchZp}
\ee
Note that in this scenario the overall coefficient $C_+$ can take any sign. 
It is worth noting that all purely leptonic meson decays such as $K_{L,S}$ or $B^0$ to $\mu \mu$ vanish in this setup since the leptonic current is vector-like \footnote{The $J=0$ constraint forces the final muon pair to be in a state with $C=+1$, whereas the vectorial current $\overline \mu \gamma \mu$ has negative $C$-parity.}. The only relevant limits then arise from $B^+ \to \pi^+ \mu \mu$, $K^+ \to \pi^+ \nu\nu$, and from LHC dimuon searches, as shown in Fig.~\ref{fig:Zplimits}.

This model also generates at the tree-level four quark operators which contribute to $\Delta F = 2$ observables:
\be
	\Delta \LL_{\Delta F = 2} = - \frac{g_q^2}{2 M_{Z^\prime}^2} \left[ (\hat n_i \hat n_j^* \overline{d_{iL}} \gamma^\alpha d_{jL})^2 + ( V_{i k} \hat n_k \hat n_l^* V^*_{j l} \; \overline{u_{iL}} \gamma^\alpha u_{jL})^2 \right]~.
\ee
For a fixed direction in quark space, $\hat n$, and a fixed value of $R_{K^{(*)}}$, we can use $\Delta F = 2$ constraints to put an upper limit on the ratio $r_{q\mu} \equiv |g_q / g_\mu|$. We can then assign a maximum value to $g_\mu$ and derive an upper limit for the $Z^\prime$ mass:
\be
	M_{Z^\prime}^{\rm lim} = \text{min}\left[ \sqrt{\frac{r_{q\mu}^{\rm lim}}{|C_+|}} |g_\mu^{\rm max}|~, ~ \sqrt{\left|\frac{g_\mu^{\rm max}g_q^{\rm max}}{C_+} \right|} \right]~,
	\label{eq:MaxZpMass}
\ee
where the first limit is from $\Delta F = 2$ observables while the second is from perturbativity. For the maximum values of the couplings we use the limits from perturbative unitarity from Ref.~\cite{DiLuzio:2017chi}, $|g_\mu^{\rm max}|^2 = 2 \pi$ and $|g_q^{\rm max}|^2 = 2\pi/3$.

\section{Theoretical expectations}
\label{sec:FlavorSymmetry}

In the previous Sections we have been agnostic about the structure of the rank one coefficients of the NP interactions, and parameterised it in terms of the unit vector $\hat n$. Here we would like to illustrate, with a flavor symmetry example, the possible theoretical expectations on the direction in flavor space at which $\hat n$ points. 

In the SM, the gauge lagrangian flavor group
\begin{equation}
\text{U}(3)^{5}\equiv\text{U}(3)_{q}\times\text{U}(3)_{\ell}\times\text{U}(3)_{u}\times\text{U}(3)_{d}\times\text{U}(3)_{e} \; 
\label{eq:U(3)5 group}
\end{equation}
is explicitly broken by the Yukawa couplings $Y_{u,d,e}$. Here, the unit vector $\hat n$, and the UV couplings from which it originates, represent an additional source of explicit breaking. In fact, we can formally assign the UV couplings introduced in the previous Section (and the SM Yukawa couplings) quantum numbers under $\text{U}(3)^{5}$:
\begin{equation}
	\SM:  \begin{cases}
		Y_u \sim 3_q \otimes \bar 3_u, \\
		Y_d \sim 3_q \otimes \bar 3_d, \\
		Y_e \sim 3_\ell \otimes \bar 3_e. \\
	\end{cases}
	\qquad 
	\qquad 
	\hat n_{i}  \propto \begin{cases}
		\beta_{i\mu}\sim \overline 3_{q}\otimes \overline{3}_{\ell} &  S_{3},\\
		\gamma_{i\mu}\sim \overline 3_{q}\otimes 3_{\ell} &  U_{1}^{\mu},\\
		M_i  \sim \overline 3_{q} &  Z',\,V' 
	\end{cases}
	\label{eq:Spurion Reps}
\end{equation}
Therefore, different models can be characterised not only in terms of the SM quantum numbers of the messengers, but also in terms of the flavor quantum numbers of the couplings. 

Correlations between the two sets of couplings in \Eq{Spurion Reps} can arise if they share a common origin. This may be the case, for example, if we assume a subgroup $\mathcal{G}\subseteq\text{U}(3)^{5}$ to be an actual symmetry of the complete UV lagrangian, and the above couplings to originate from its spontaneous breaking by means of a common set of ``flavon'' fields. 

Correlations cannot arise if $\mathcal{G}$ coincides with the full $\text{U}(3)^{5}$. The quantum numbers of the relevant flavons coincide in this case with the transformation properties in \Eq{Spurion Reps}. Therefore, the flavons entering the Yukawas and the NP couplings are in this case entirely independent. In particular, the ROFV assumption is not compatible with the Minimal Flavor Violation one~\cite{DAmbrosio:2002vsn}. We therefore need to consider proper subgroups of $\text{U}(3)^{5}$. Among the many possibilities, let us consider the $\mathcal{G} = \text{U}(2)^{5}$ subgroup of transformations on the first two fermion families. The latter extends the quark $\text{U}(2)^{3}$~\cite{Barbieri:2011ci} to the leptons, relevant for two of the three NP couplings in \Eq{Spurion Reps}. Some of the conclusions we will draw hold for a generic extension to $\text{U}(2)^{3}\times \mathcal{G}_l$, where $\mathcal{G}_l$ only acts on leptons. 

The fact that correlations can arise in the U(2) case is not a surprise. In the unbroken limit, the versor $\hat n$ and the SM Yukawas must leave the same $\U(2)_q$ subgroup invariant, and are therefore aligned (although no flavor violation would be generated in such a limit). In order to investigate them, we write all the $\mathcal{G}$-violating couplings as VEVs of flavons with irreducible $\mathcal{G}$ quantum numbers, and assume the UV theory to contain at most one flavon of each type. The predictions that follow then depend on the structure of the flavon sector, as we now discuss.

Let us first consider the case, which we will refer to as ``minimally broken'' $\U(2)^5$, in which no flavon is charged under both the quark $\U(2)^3$ and the lepton $\U(2)^2$, or $\mathcal{G}_l$.
In such a case one finds a precise correlation between the first two components of the unit vector $\hat n$ and of the third line of the CKM matrix: $\hat n_1/\hat n_2 = V^*_{td}/V^*_{ts}$, up to corrections of order $m_s/m_b$.
We then have
\be
	\hat n_{U 2} \propto \left( c_{U2} e^{i \gamma} V_{td}^*, ~ c_{U2} e^{i \gamma} V_{ts}^*, ~1\right)~,
	\label{eq:U2direction}
\ee
where $c_{U2}\sim \mathcal{O}(1)$ and the normalisation is fixed by the condition $| \hat n |^2 = 1$.
Comparing with the parametrization in Eq.~\eqref{eq:VersorDef}, one gets:
\begin{equation}
	\tan \phi = \frac{|V_{ts}|}{|V_{td}|} \;, \quad
	\tan \theta \approx c_{U2} |V_{ts}| \;, \quad
	\alpha_{bd} = - \arg(V_{td}) + \gamma \;, \quad
	\alpha_{bs} = - \arg(V_{ts}) + \gamma \;. 
\label{eq:U2minimalpar}
\end{equation}
This prediction also applies in the case of $Z'$ or $V'$ messengers, independent on whether $\U(2)^5$ is minimally broken or not. 

\begin{figure}[t]
\centering
\includegraphics[width=0.48\hsize]{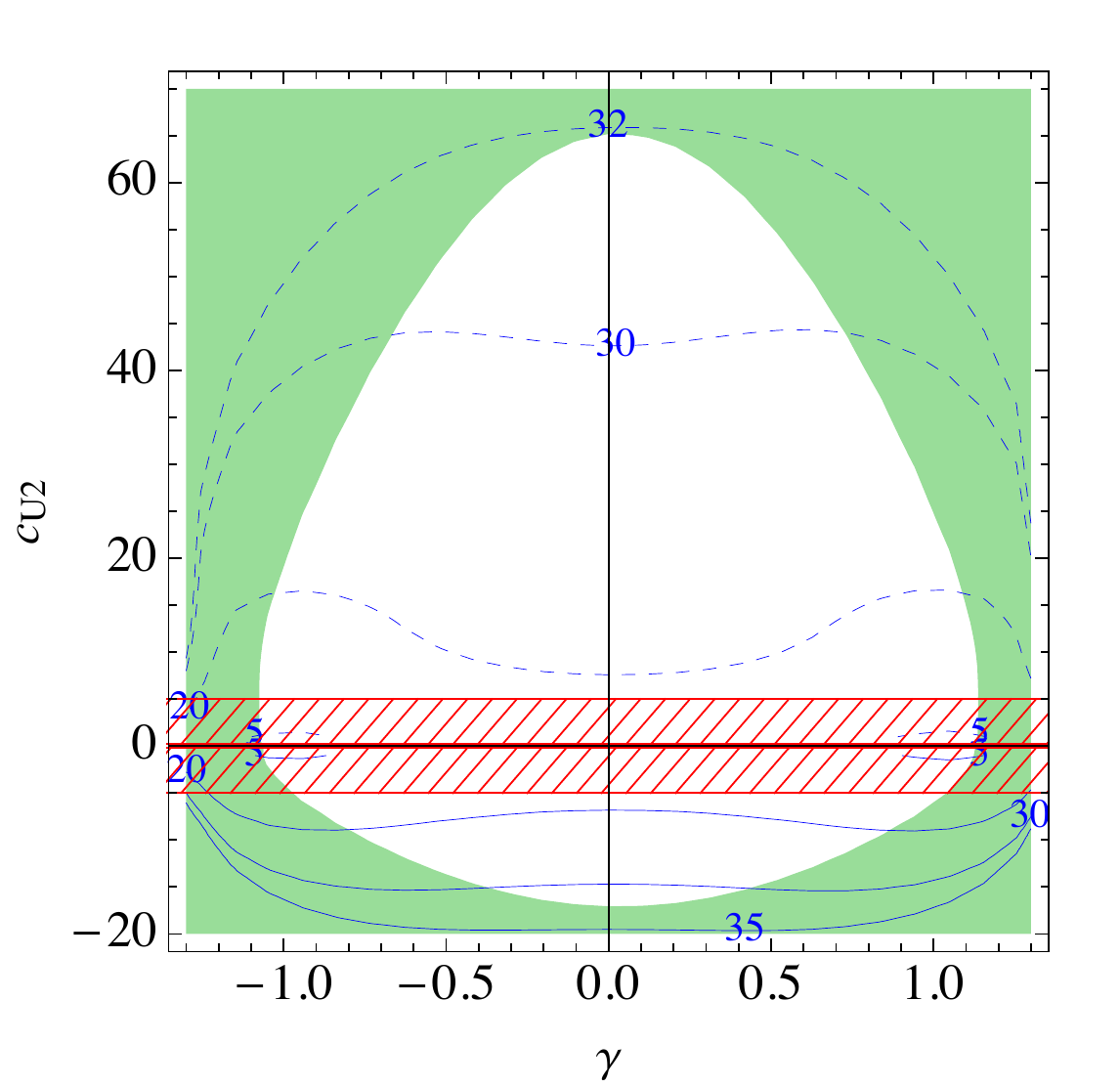}
\caption{Region excluded at $95\%$CL in a global fit of $bs\mu\mu$ clean observables and the other $d_i d_j \mu\mu$ ones listed in \Tab{Correlated-observables}, assuming the structure of Eq.~\eqref{eq:U2direction}, in the plane $(\gamma, c_{U2})$. The relevant observable in the excluded region is $K_L \to \mu^+ \mu^-$. The range $|c_{U2}| \in [0.2 - 5]$ is highlighted as the meshed red region. Blue lines indicate the size of the overall coefficient $|C_L|^{-1/2}$ (in TeV), as extracted from the fit. Solid (dashed) lines are for positive (negative) values of $C_L$.
}\label{fig:minimalU2}
\end{figure}

In the presence of flavons charged both under the quark and lepton part of the flavor group, or in the presence of more flavons with the same flavor quantum numbers, the above prediction does not need to hold. On the other hand, with reasonable assumptions on the size of the flavons, one obtains a generic correlation between the unit vector $\hat n$ and the third line of the CKM matrix, which holds up to $\ord{1}$ factors: $\hat n = (\mathcal{O}(V_{td}), \mathcal{O}(V_{ts}), \mathcal{O}(1))$. We can parametrize such a scenario in full generality as
\be
	\hat n \propto \left(  a_{bd} e^{i \alpha_{bd}} |V_{td}|, ~ a_{bs} e^{i \alpha_{bs}} |V_{ts}|, ~ 1\right) ,
	\label{eq:TopLikeDirection}
\ee
where $a_{bd}$ and $a_{bs}$ are $\mathcal{O}(1)$ real parameters. The area in the $(\phi, \theta)$ plane corresponding to values $|a_{bs, bd}| \in [0.2 - 5]$ is shown as a meshed-red one in the plots of Figs.~\ref{fig:Minimal},\ref{fig:S3limits},\ref{fig:U1limits}. 

The correlation in \Eq{TopLikeDirection} is also found with different flavor groups, and in models with partial compositeness (and no flavor group). In the limit in which the top quark exchange dominates FCNC processes, the SM itself satisfies the ROFV condition, with $\hat n = (V^*_{td},V^*_{ts},V^*_{tb})$ also in the form above.

One comment is in order about the role of the lepton flavor sector. The latter can play a twofold role. On the one hand, it can affect the prediction for the direction of $\hat n$. This can be the case for $S_3$ and $U^\mu_1$ messengers, for which $\hat n$ is associated to the muon row of the $\beta$ and $\gamma$ matrices in \Eq{Spurion Reps}. On the other hand, the lepton flavor breaking can affect the overall size of the effect. The anomalies require in fact a breaking of $\mu$-$e$ lepton universality, whose size is associated to the size of $\U(2)_l$ breaking. A sizeable breaking is necessary in order to account for a NP effect as large as suggested by the $B$-meson anomalies. A detailed analysis of the implications of the anomalies for lepton flavor breaking and for processes involving other lepton families is outside the scope of this work. 

We now focus on the case of minimally broken $\U(2)^5$, \Eq{U2minimalpar}. The 95\%CL limit in the plane $(\gamma, c_{U2})$, from our global fit of $bs\mu\mu$ clean observables (see App.~\ref{app:bsmumu}) and the other $d_i d_j \mu\mu$ ones (\Tab{Correlated-observables}) is shown in Fig.~\ref{fig:minimalU2}-Left. The relevant observable in the excluded region is $K_L \to \mu^+ \mu^-$.
For positive (negative) values of $C_+$ we obtain a limit $c_{U2} \gtrsim -20$ ($\lesssim 65$), which are well outside the natural region predicted by the flavor symmetry.

Another interesting point is that, within the parametrization \eqref{eq:U2direction}, one has 
\be
	R_K\approx R_\pi
	\label{eq:RKRpiSU2}
\ee
up to $\mathcal{O}(m_s/m_b)$ corrections, where: 
\be 
	R_H \equiv \dfrac{\text{Br}(B\to H \mu ^+ \mu ^-)_{[1,6]}}{\text{Br}(B\to H e ^+ e ^-)_{[1,6]}} \qquad (H=K,\pi),
\label{eq:LFUratios}
\ee
so that a comparison between the two observables could in principle rule out, in this context, the $V'$ and $Z'$ cases, as well as minimally broken $\text{U}(2)^5$. 
Assuming no NP in the electron channels, we also have\footnote{Using the LFU ratios \eqref{eq:LFUratios} is of course advisable from the theoretical point of view. Unfortunately, there are no measurements of $\text{Br}(B\to \pi e ^+ e ^-)$, at present.}:
\be 
	R_\pi = \dfrac{\text{Br}(B\to \pi \mu ^+ \mu ^-)_{[1,6]}}{\text{Br}(B\to \pi \mu ^+ \mu ^-)_{[1,6]}^{\text{SM}}}.
\label{eq:RpiAssumption}
\ee 
The RHS of Eq. \eqref{eq:RpiAssumption} is, experimentally, (cf. Table \ref{tab:Correlated-observables}):
\be
\dfrac{\text{Br}(B\to \pi \mu ^+ \mu ^-)_{[1,6]}^{\text{exp}}}{\text{Br}(B\to \pi \mu ^+ \mu ^-)_{[1,6]}^{\text{SM}}}=0.70\pm 0.30,
\label{eq:RpiExpt}
\ee
showing no tension neither with the SM prediction, nor with the $\text{U(2)}^5$ prediction \eqref{eq:RKRpiSU2}.
Another prediction of this setup is for the branching ratio of $B^0_d \to \mu^+ \mu^-$ with respect to $B_s \to \mu^+ \mu^-$:
\be
	\dfrac{\Br(B^0_s \to \mu ^+ \mu ^-)}{\Br(B^0_s \to \mu ^+ \mu ^-)^{\text{SM}}} \approx \dfrac{\Br(B^0_d \to \mu ^+ \mu ^-)}{\Br(B^0_d \to \mu ^+ \mu ^-)^{\text{SM}}}~.
	\label{eq:B0BsmumuSU2}
\ee
The two predictions \eqref{eq:RKRpiSU2} and \eqref{eq:B0BsmumuSU2} are independent on the specific chiral structure of the muon current. If the operators responsible for $R_{K^{(*)}}$ are left-handed only, the two ratios in Eq.~\eqref{eq:B0BsmumuSU2} are also predicted to be of the same size as $R_K$ and $R_\pi$, up to $\mathcal{O}(2\%)$ corrections.
Such corrections are however negligible when compared to the expected precision in the measurements of these relations, which is at best of $\approx 4\%$, c.f. \Tab{Prospects}. 
It is perhaps worth pointing out that the predictions in Eqs.~(\ref{eq:RKRpiSU2},\ref{eq:B0BsmumuSU2}) are a consequence of the minimally broken $\U(2)^5$ flavor symmetry alone, independently of the ROFV assumption. This can be understood from the fact that the $b-s$ and $b-d$ transitions are related by $\U(2)^5$ symmetry as $C_{S,T}^{bd}/C_{S,T}^{bs} = (V_q)^1/(V_q)^2 = V_{td}^*/V_{ts}^*=n^1_{U2}/n^2_{U2}$, where $V_q$ is the spurion doublet under $\U(2)_q$.

\section{Future Prospects}
\label{sec:Prospects}

\begin{table}[t]
\centering
\begin{tabular}{|c|c l c|} \hline
Observable & Expected sensitivity & Experiment  & Reference \\
\hline
\hline
\multirow{2}{*}{$R_K$} & $0.7 \; (1.7) \%$ &  LHCb 300 (50) fb$^{-1}$  & \cite{Bediaga:2018lhg} \\
 & $3.6 \; (11) \%$ &  Belle II 50 (5) ab$^{-1}$  & \cite{Kou:2018nap} \\
\hline
\multirow{2}{*}{$R_{K^{*}}$} & $0.8 \; (2.0) \%$ &  LHCb 300 (50) fb$^{-1}$  & \cite{Bediaga:2018lhg} \\
 & $3.2 \; (10) \%$ &  Belle II 50 (5) ab$^{-1}$  & \cite{Kou:2018nap} \\
\hline
$R_\pi$ & $4.7 \; (11.7) \%$ &  LHCb 300 (50) fb$^{-1}$  & \cite{Bediaga:2018lhg} \\
\hline
\multirow{2}{*}{$\Br(B_s^0 \to \mu^+ \mu^-)$}  & $4.4 \; (8.2)\%$ & LHCb 300 (23) fb$^{-1}$ & \cite{Bediaga:2018lhg,Cerri:2018ypt} \\
 & $7 \; (12)\%$ & CMS 3 (0.3) ab$^{-1}$ & \cite{Cerri:2018ypt} \\
\hline
\multirow{2}{*}{$\Br(B^0_d \to \mu^+ \mu^-)$} & $9.4 \; (33) \%$ & LHCb 300 (23) fb$^{-1}$ & \cite{Bediaga:2018lhg,Cerri:2018ypt} \\
 & $16 \; (46) \%$ & CMS 3 (0.3) ab$^{-1}$ & \cite{Cerri:2018ypt} \\
\hline 
$\Br(K_S \to \mu^+ \mu^-) $ & $\sim 10^{-11}$ & LHCb 300fb$^{-1}$ & \cite{Bediaga:2018lhg,Cerri:2018ypt} \\
\hline
\multirow{3}{*}{$\Br(K_L \to \pi^0 \nu\nu)$} & $\sim 1.8 \times 10^{-10}$ & KOTO phase-I $^7$ &  \\\
 & $20\%$ & KOTO phase-II $^7$ &  \\
 & $20\%$ & KLEVER & \cite{Ambrosino:2019qvz} \\
\hline
$\Br(K^+ \to \pi^+ \nu\nu) $ & $10\%$ & NA62 goal & \cite{Ruggiero:2017hjh} \\
\hline
\end{tabular}
\caption{\label{tab:Prospects} Future prospects for the precision reach in various flavor observables. The expected sensitivity in percent are quoted with respect to the Standard Model prediction.} 
\end{table}

\begin{figure}[t]
\centering
\includegraphics[width=0.49\hsize]{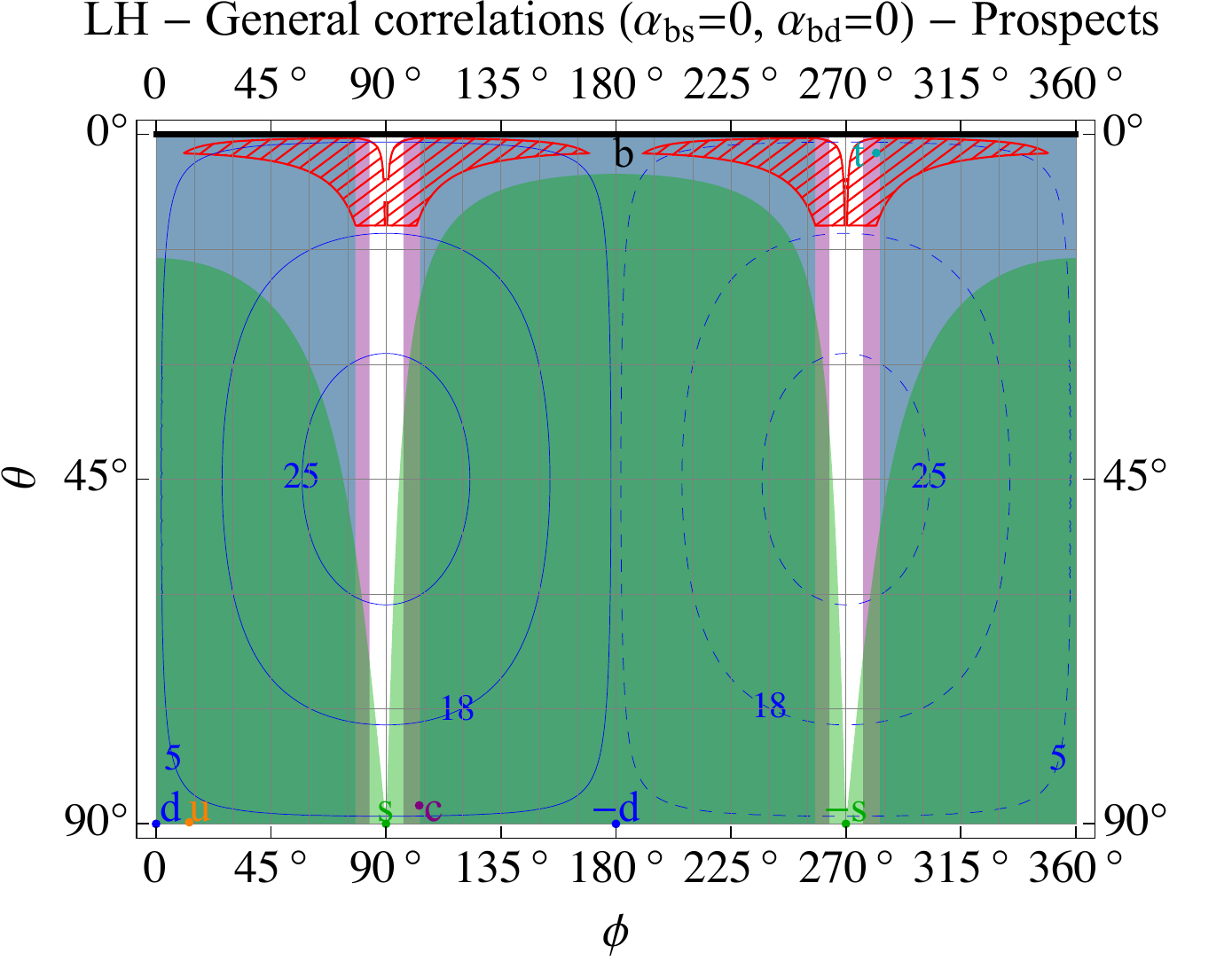}
\includegraphics[width=0.49\hsize]{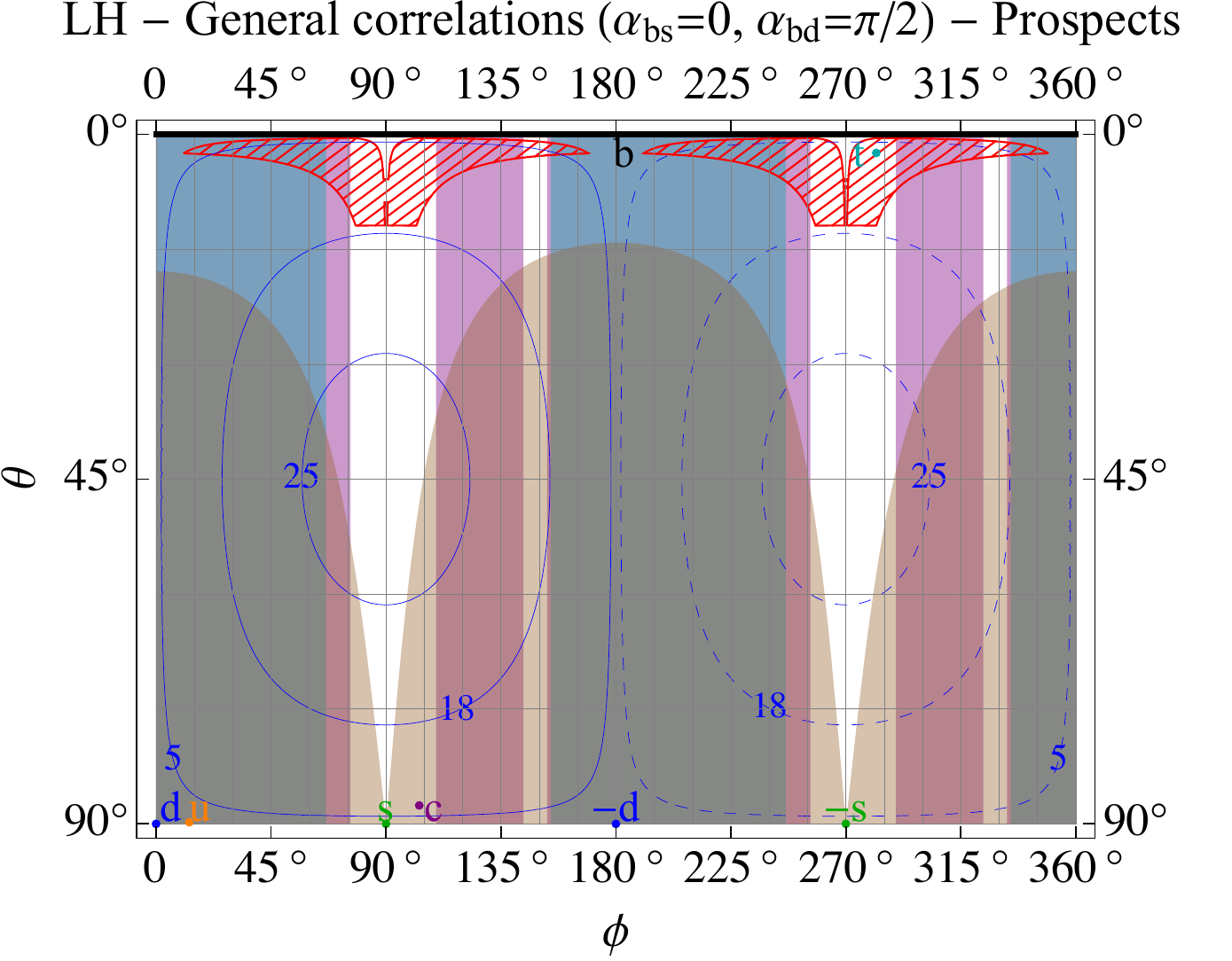}\\
\includegraphics[width=0.52\hsize]{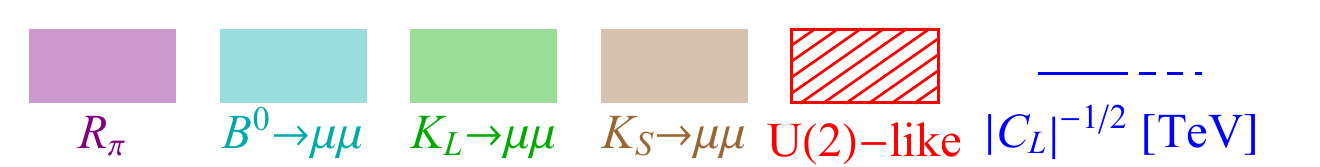}
\caption{Future prospects for the exclusion limits in the plane $(\phi, \theta)$ for two choices of the phases $\alpha_{bs}$ and $\alpha_{bd}$ from observables with direct correlation with $R_{K^{(*)}}$. For $K_L \to \mu\mu$ we use the present bound.}
\label{fig:Minimal_prospect}
\end{figure}

Future measurements by LHCb, Belle-II, and other experiments are expected to improve substantially the precision of most of the observables studied in the present work. We collect in Table~\ref{tab:Prospects} the relevant prospects. 

First of all, the anomalous observables themselves, $R_K$ and $R_{K^{*}}$, are expected to be tested with sub-percent accuracy by LHCb with 300fb$^{-1}$ of luminosity. Furthermore, a larger set of observables sensitive to the same partonic transition $b \to s \mu^+ \mu^-$ will be measured (such as $R_\phi$, $R_{pK}$ and $Q_5$ for example \cite{Bediaga:2018lhg}). This will allow to confirm or disprove the present anomalies and to pinpoint the size of the New Physics contribution with high accuracy.

The leptonic decays $B^0_{(d,s)}\to \mu ^+ \mu ^-$ will be crucial for discriminating between the $\mathcal{O}_9$ and $\mathcal{O}_L$ scenarios. As to the $B\to \pi \ell^+ \ell ^-$ channels, we note that the power of the muon-specific $\text{Br}(B^+\to \pi ^+ \mu^+ \mu ^-)$ as a probe of NP is, already at present, limited by theoretical uncertainties \cite{Khodjamirian:2017fxg}. The situation improves substantially for the LFU ratio $R_\pi$ (cf. Eq. \eqref{eq:LFUratios}),  for which, as already noted, $\U(2)^5$ flavor symmetry predicts $R_\pi = R_K$ and for which LHCb is expected to reach a $\sim 4.7\%$ sensitivity with 300 fb$^{-1}$ of luminosity \cite{Bediaga:2018lhg}.
As can be seen in Fig.~\ref{fig:Minimal_prospect}, these channels will be able to cover almost the complete parameter space of the setup studied here, particularly if the phase $\alpha_{bd}$ is small.

\begin{figure}[t]
\centering
\includegraphics[width=0.45\hsize]{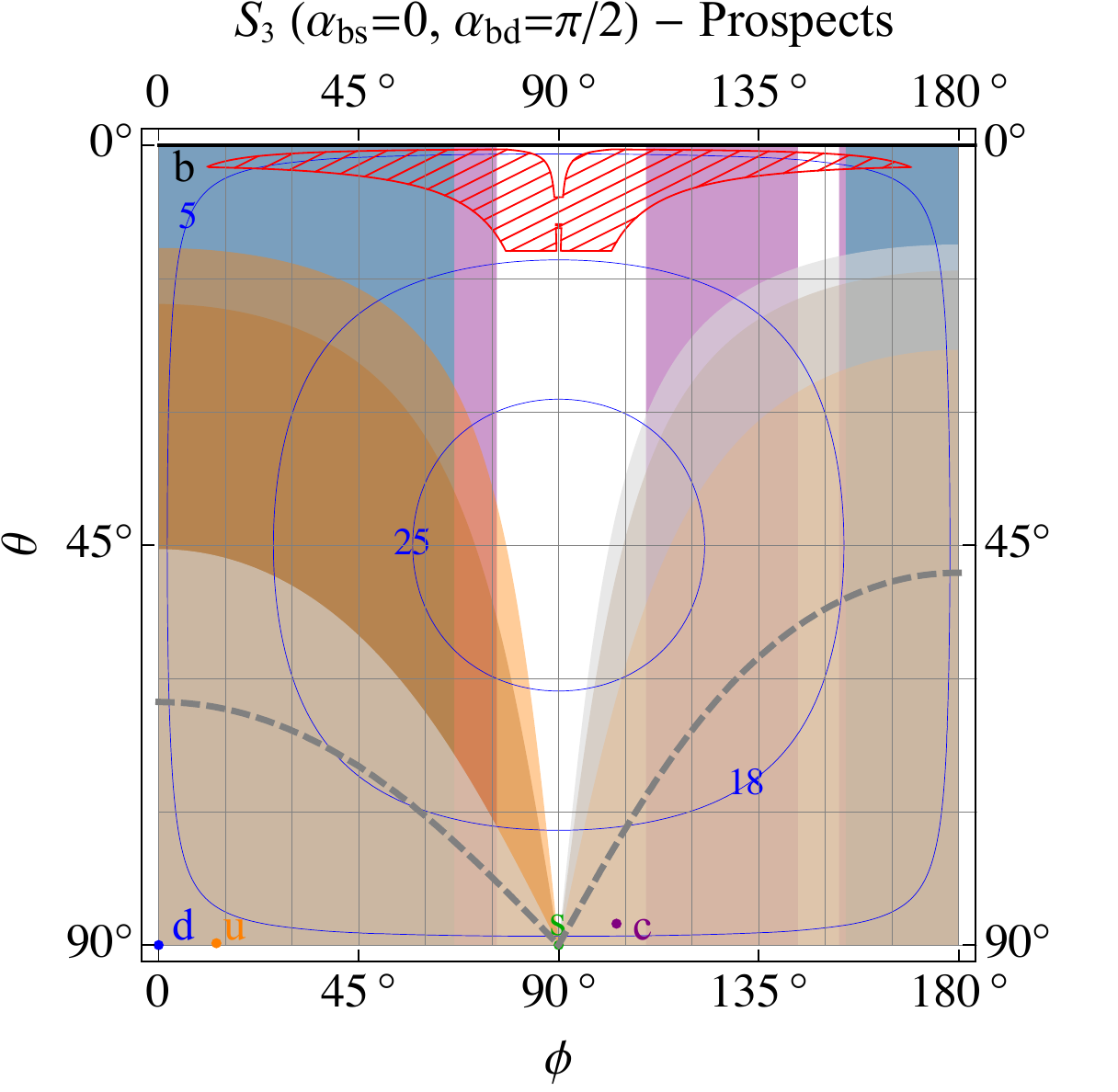}\!\!\!\!\!
\includegraphics[width=0.55\hsize]{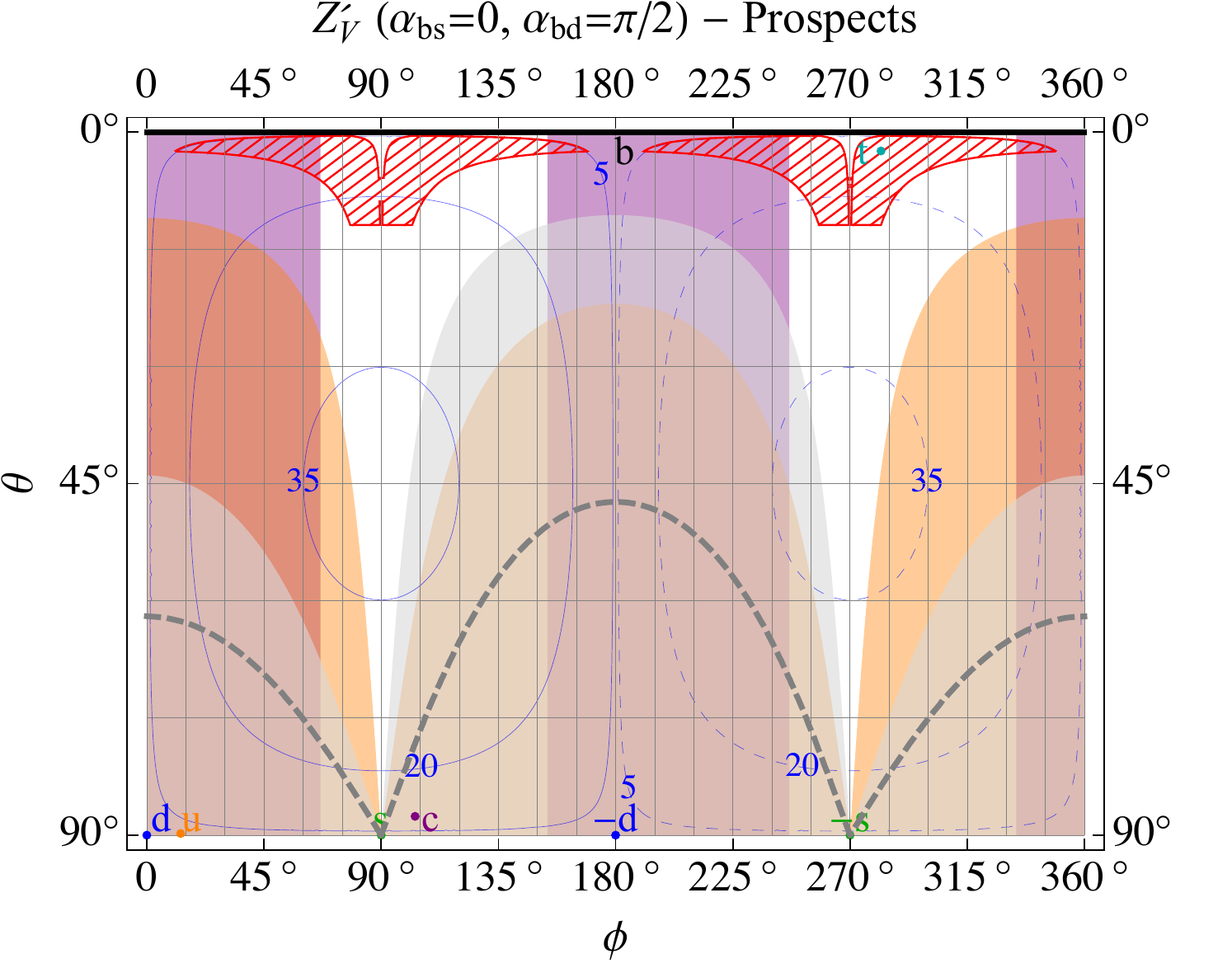}\\
\includegraphics[width=0.6\hsize]{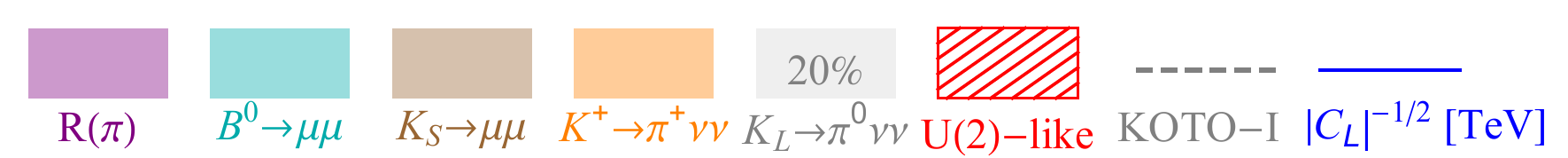}
\caption{Future prospects for the exclusion limits in the $(\phi, \theta)$ plane for the scalar leptoquark $S_3$ (left) and the vector singlet $Z^\prime$ with vector-like couplings to muons (right), for a choice of phases. The orange and gray regions correspond to the future expected limits from $K^+ \to \pi^+ \nu\nu$ (NA62 \cite{Ruggiero:2017hjh}) and $K_L \to \pi^0 \nu\nu$ (KLEVER \cite{Ambrosino:2019qvz} or KOTO phase-II), respectively. The dashed gray line corresponds to the expected limit on $K_L \to \pi^0 \nu\nu$ from KOTO phase-I.}\label{fig:prospectsKpinunu}
\end{figure}

In all cases where $C_S \neq C_T$, such as in the $S_3$ and $Z^\prime$ models, other relevant channels which will improve substantially in sensitivity are $\Br(K^+ \to \pi^+ \bar\nu\nu)$ and $\Br(K_L \to \pi^0 \bar\nu\nu)$.
The former is expected to be measured with a 10\% accuracy by NA62 \cite{Ruggiero:2017hjh} in the next few years, while, for the latter, the KOTO experiment at JPARC \cite{Ahn:2018mvc} should reach
a single-event sensitivity at the level of the SM branching ratio, with a signal to background ratio $\sim 1$, which translates to a projected $95\%$CL limit of $\sim 5.4$ times the SM value, i.e. $\sim 1.8 \times 10^{-10}$. A possible future upgrade of the whole KOTO experiment (stage-II)\footnote{From a CERN EP seminar given in February 2019 (slides at \url{https://indico.cern.ch/event/799787/attachments/1797668/2939627/EPSeminar_YuChen.pdf}) and of a talk presented at the Rencontres de Moriond 2019 (slides at \url{http://moriond.in2p3.fr/2019/EW/slides/1_Sunday/1_morning/5_Nanjo.pdf}).}, or the proposed KLEVER experiment at CERN SPS \cite{Ambrosino:2019qvz}, could both reach a $\sim 20\%$ sensitivity of the $\Br(K_L \to \pi^0 \bar\nu\nu)$ SM value.
An example for the prospects due to these observables for a particular choice of phases in the two simplified models are shown in Fig.~\ref{fig:prospectsKpinunu}.

\section{Conclusions}
\label{sec:Conclusions}
If the flavor anomalies in $b \to s \mu \mu$ transitions are experimentally confirmed, they will provide important information about the flavor structure of the underlying New Physics. The latter can be tested by studying possible correlations with other measurements in flavor physics. 
In this work we assumed that the putative NP, responsible for the anomalous effects,
couples to SM left-handed down quarks in such a way to generate a rank-one structure in the novel flavor-violating sector. We dub such a scenario \emph{Rank-One Flavor Violation} (ROFV). Such a structure can result from a number of well motivated UV completions for the explanation of the flavor anomalies, in which a single linear combination of SM quark doublets couples to the relevant NP sector. This automatically includes all single-leptoquark models, and models where LH quarks mix with a single vector-like fermion partner. As these examples reveal, the ROFV condition might not originate from symmetry but rather as a feature of the UV dynamics.

Varying the direction associated to the NP ($\hat{n}$) in $\U(3)_q$ flavor space, we identified the most important observables that can be correlated to the flavor anomalies. The more model-independent correlations are with $d_i \to d_j \mu \mu$ transitions (and their crossed symmetric processes). A large part of the parameter space is probed by the measurement of the branching ratio of $B^+ \to \pi^+ \mu \mu$. While the sensitivity to NP effects in this channel is limited by the large hadronic uncertainty of the SM prediction, future measurements of the theoretically clean ratio $R_\pi$ are going to provide further information on $b \to d$ flavor violations. 
Among the transitions involving the first two generations of quarks ($s \to d$), the $K_L \to \mu \mu$ decay rate has a major impact and it is particularly sensitive to the phases of our parametrization. Unfortunately, future prospects in this channel are limited by theory uncertainties in the SM prediction of the long-distance contribution to the decay. A sizeable improvement by LHCb is instead expected in the limit on the $K_S \to \mu\mu$ decay rate.

While the former conclusions rely only on our rank-one hypothesis, more model dependent correlations can be established once the relevant effective operators are embedded into the SMEFT or in the presence of specific mediators. An example of such correlations is given by $d_i \to d_j \nu \nu$ processes, and we have in fact shown that present data from $K^+ \to \pi^+ \nu\nu$ are particularly relevant to the leptoquark $S_3$ or vector $Z'$ simplified models. 

From a more theoretical point of view, we investigated whether the flavor violation associated to the NP can be connected to the one present in the SM Yukawa sector. A generic expectation is that the leading source of $\U(3)_q$ breaking in the NP couplings is provided by a direction in flavor space close to the one identified by the top quark. Indeed, we showed in a concrete example based on a flavor symmetry that the vector $\hat{n}$ turns out to be correlated to the third line of the CKM matrix, as in Eq.~(\ref{eq:TopLikeDirection}). Remarkably, a large portion of the theoretically favoured region (red meshed lines region in our plots) survive the bounds from current flavor physics data. Our order of magnitude predictions can be narrowed down under further theoretical assumptions. For example a minimally broken $\U(2)^5$ flavor symmetry predicts $R_K = R_\pi$ and the ratio $\Br (B_s \to \mu \mu) / \Br (B_d \to \mu \mu) $ to be SM-like (up to small corrections of a few percent).

In our last section we explored future prospects for the exclusion limits in the ROFV framework. In the near future a series of experiments will be able to cover almost all of the parameter space identified by our ansatz. Indeed, in the next few years, significant information will be provided by the NA62 and KOTO experiments, thanks to precise measurements of the $K^+ \to \pi^+ \nu\nu$ and $K_L \to \pi^0 \nu\nu$ decays, while on a longer time scale results from LHCb and Belle II will almost completely cover our parameter space (and test the minimally broken $\U(2)^5$ model).

A confirmation of the $B$-meson anomalies would open a new era in high energy physics. In this enticing scenario, studying correlations of the anomalies to other observables would provide a powerful mean to investigate the flavor structure of New Physics.

\subsection*{Acknowledgments}
We thank Dario Buttazzo and Joe Davighi for useful discussions. 
The authors are grateful to the Mainz Institute for Theoretical Physics (MITP) for its hospitality and its partial support during the completion of part of this work. AR acknowledges partial support by the PRIN project ``Search for the Fundamental Laws and Constituents'' (2015P5SBHT\_002) and the FP10 ITN projects ELUSIVES 
(H2020-MSCA-ITN-2015-674896) and INVISIBLES-PLUS (H2020-MSCA-RISE-2015-690575). DM and MN acknowledge partial support by the INFN grant SESAMO.

\appendix

\section{Simplified fit of clean $b s \mu \mu$ observables}
\label{app:bsmumu}

In this section we focus on the \emph{clean} observables sensible to the $b s \mu \mu$ local interactions. These are defined as those for which the Standard Model prediction is free of large uncertainties, typically due to a poor knowledge of the non-perturbative QCD dynamics.
The relevant observables are the Lepton Flavor Universality (LFU) ratios $R_K$ and $R_{K^{*}}$, in the specific bins of $q^2$ measured experimentally, as well as the branching ratio of $B_s \to \mu^+ \mu^-$, which can be predicted with good accuracy \cite{Bobeth:2013uxa,Bordone:2016gaq}. The experimental measurements for these observables are collected in \Tab{bsmumuObs}. We included also the latest results for $R_K$ which combines data from 2011 to 2016 \cite{Aaij:2019aa}, as well as the analysies on $R_{K^{*0}}$ and $R_{K^{*+}}$ presented by Belle at the Rencontres de Moriond 2019 \cite{2019:BelleRKst}.\footnote{The slides can be found at the following url: \url{http://moriond.in2p3.fr/2019/EW/slides/6_Friday/1_morning/1_Markus_Prim.pdf}.} For updated global fits including all relevant observables we refer to \cite{DAmico:2017mtc,Alguero:2019aa,Alok:2019ufo,Ciuchini:2019usw,Aebischer:2019mlg}.

\begin{figure}[t]
\centering
\includegraphics[width=0.35\hsize]{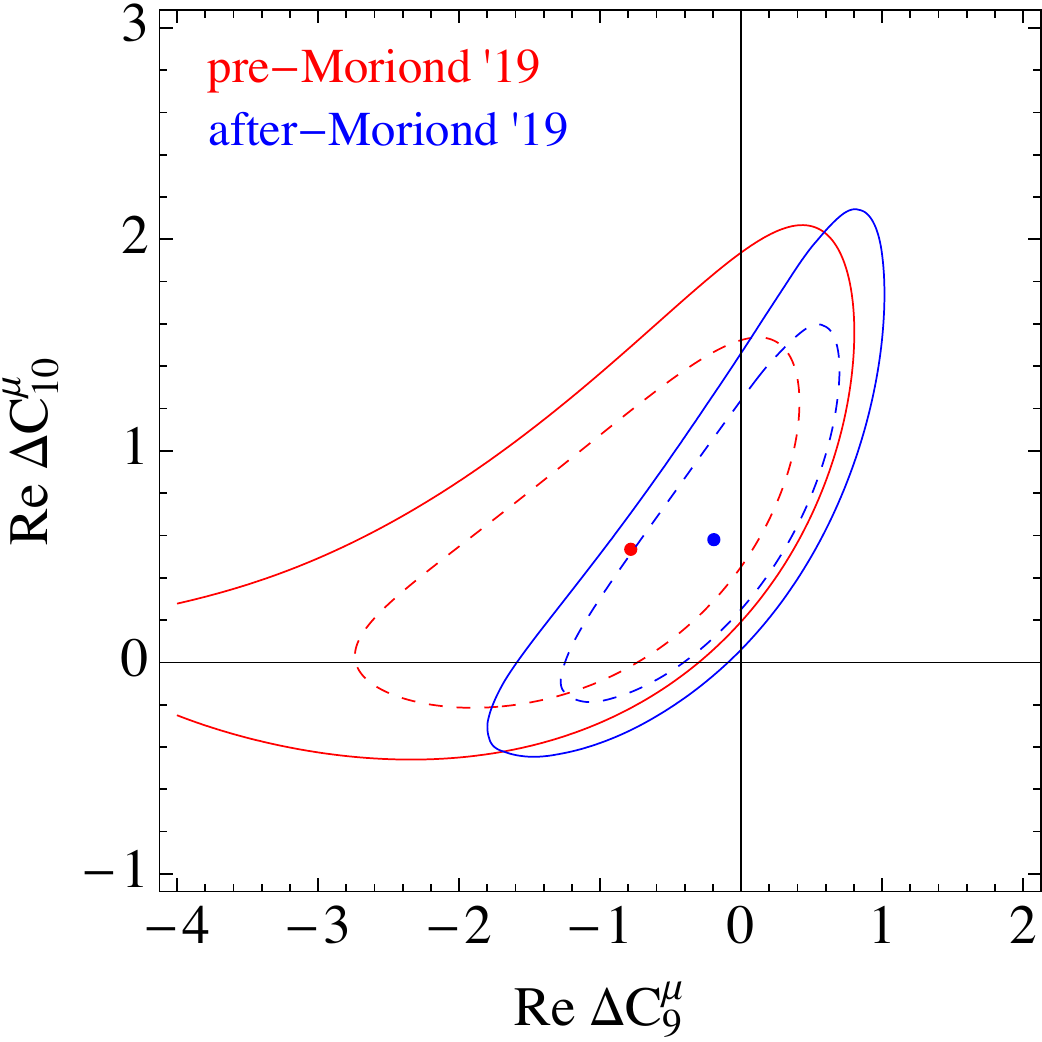}
\includegraphics[width=0.35\hsize]{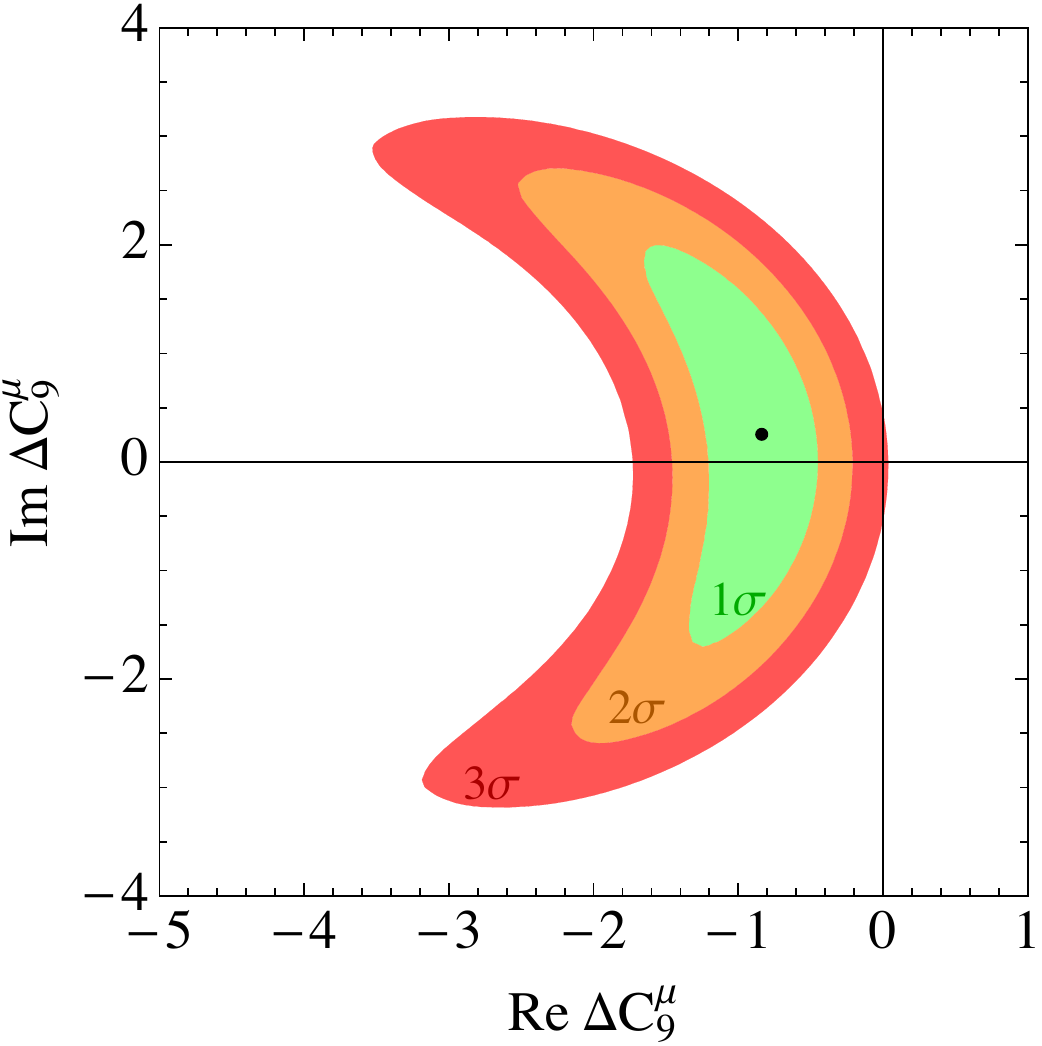}\\
\includegraphics[width=0.35\hsize]{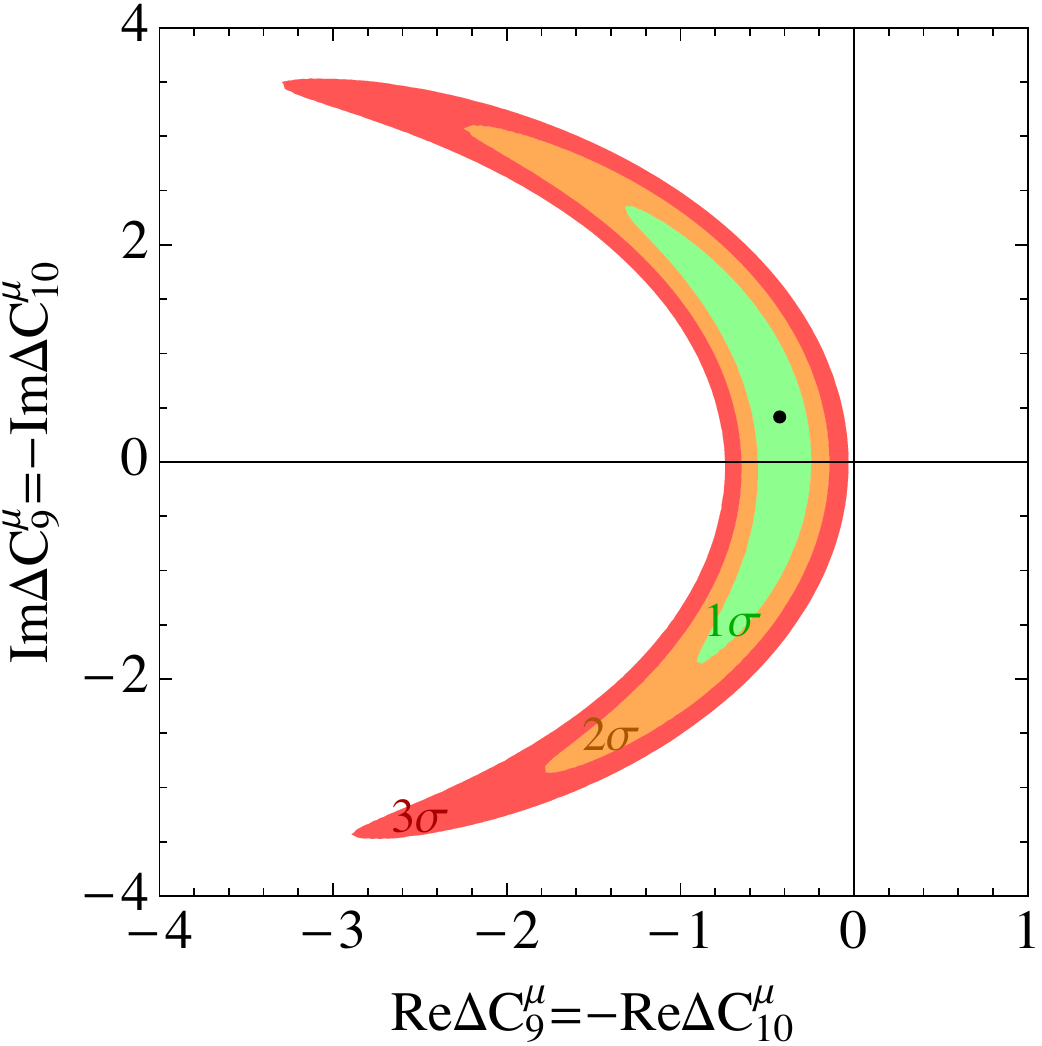}
\includegraphics[width=0.35\hsize]{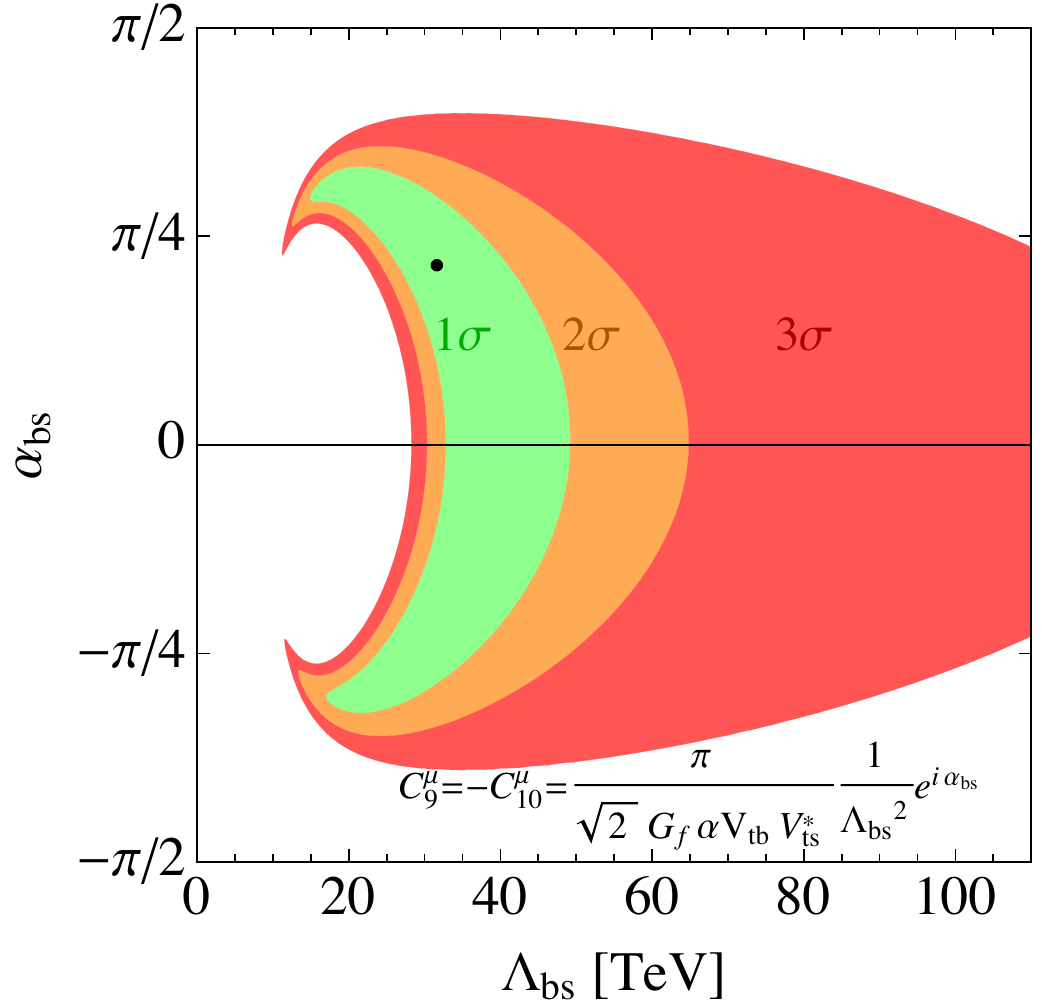}
\caption{2D fits of the \emph{clean} $bs\mu\mu$ observables listed in \Tab{bsmumuObs}. In the top-left panel the dots represent the best-fit points while dashed (solid) lines are $95\%$CL ($99\%$CL) contours. In the other panels the black dot is the best-fit point and the green, orange, and red regions are such that $\Delta \chi^2 \leq$ 2.28 ($68\%$CL), 5.99 ($95\%$CL), and 11.6 ($99\%$CL), respectively. }\label{fig:bsmumuFit}
\end{figure}

We are interested in New Physics operators with current-current structure and left-handed quarks, since they allow the best fits to the observed experimental anomalies:
\be
	\mathcal{L}_{\rm eff}^{\rm NP} \supset \frac{G_F \alpha}{\sqrt{2} \pi} V_{tb} V_{ts}^* \left[ \Delta C_9^\mu (\bar s \gamma^\mu P_L b) (\bar \mu \gamma_\mu \mu) + \Delta C_{10}^\mu (\bar s \gamma^\mu P_L b) (\bar \mu \gamma_\mu \gamma_5 \mu) \right] + h.c. ~.
\ee

We derive the dependence of the clean observables on the $\Delta C_9^\mu$ and $\Delta C_{10}^\mu$ coefficients using the  expressions of the differential rates and form factors from Refs.~\cite{Altmannshofer:2008dz,Bobeth:2011nj,Becirevic:2012fy,Straub:2015ica}:
\bea
	R_K([1.1-6]) \!\!&\approx&\!\! 1.00 + 0.24 (\Re \Delta C_{9}^\mu - \Re \Delta C_{10}^\mu) + 0.029 (|\Delta C_{9}^{\mu}|^{2} + |\Delta C_{10}^{\mu}|^{2}),~ \nonumber \\
	R_{K^{*}}([0.045-1.1]) \!\!&\approx&\!\! 0.93 + 0.057 \Re \Delta C_{9}^\mu -  0.10 \Re \Delta C_{10}^\mu  + 0.012 (|\Delta C_{9}^{\mu}|^{2} + |\Delta C_{10}^{\mu}|^{2}),\nonumber \\
	R_{K^{*}}([1.1-6]) \!\!&\approx&\!\! 1.00 + 0.21 \Re \Delta C_{9}^\mu - 0.29 \Re \Delta C_{10}^\mu  + 0.035 (|\Delta C_{9}^{\mu}|^{2} + |\Delta C_{10}^{\mu}|^{2}), \nonumber \\
	R_{K^{*}}([15-19]) \!\!&\approx&\!\! 1.00 + 0.24 \Re \Delta C_{9}^\mu - 0.25 \Re \Delta C_{10}^\mu + 0.030 (|\Delta C_{9}^{\mu}|^{2} + |\Delta C_{10}^{\mu}|^{2}), \nonumber \\
	\frac{\Br(B_s^0 \to \mu^+ \mu^-)}{\Br(B_s^0 \to \mu^+ \mu^-)_\SM} \!\!&\approx&\!\! \left| 1 + \frac{\Delta C_{10}^\mu}{C_{10, \text{ eff}}^\mu} \right|^2~,
\eea
where $C_{10, \text{ eff}}^\mu = - 4.103$ \cite{Altmannshofer:2008dz} describes the short-distance SM contribution. These are in good agreement with the numerical expressions of Ref.~\cite{Geng:2017svp}. In the above expression we fixed the central value for the coefficients of the form factor parametrization. For $B_s^0 \to \mu\mu$ we combine the LHCb \cite{Aaij:2017vad} and ATLAS \cite{Aaboud:2018mst} measurements assuming Gaussian distributions, and as the SM prediction we take $\Br(B_s^0 \to \mu\mu)_\SM = (3.65 \pm 0.23) \times 10^{-9}$ \cite{Bobeth:2013uxa}.

We perform a simple $\chi^2$ fit of these observables for a set of assumptions on the NP coefficients, the results are shown in \Fig{bsmumuFit}. In the top-left panel we show a comparison with the fit performed with or without the latest results presented at the Rencontres du Moriond 2019 conference, in the $(\Re \Delta C_9^\mu, \Re \Delta C_{10}^\mu)$ plane. We are particularly interested in the case where the NP coefficients has a non-vanishing complex phase. Our results, when removing the latest results presented at Moriond 2019, are in good agreement with Ref.~\cite{Altmannshofer:2014rta}, which also shows a fit including imaginary parts for the NP coefficients.

The lower-right panel of \Fig{bsmumuFit} shows the fit in the parametrization of the left-left operator we mostly focus on in this work:
\be
	\LL_{\rm eff} \supset \frac{e^{i \alpha_{bs}}}{\Lambda_{bs}^2} (\bar s_L \gamma^\mu b_L) (\bar \mu_L \gamma_\mu \mu_L ) + h.c.~,
\ee
which is related to the standard parametrization by
\be
	\frac{e^{i \alpha_{bs}}}{\Lambda_{bs}^2} = \frac{G_F \alpha}{\sqrt{2} \pi} V_{tb} V_{ts}^*  (\Delta C_9^\mu - \Delta C_{10}^\mu)~.
\ee
The result of the fit in this parametrization can be seen in the bottom-right panel of Fig.~\ref{fig:bsmumuFit}. The best-fit point is found for $\Lambda_{bs} \approx 31.6 \TeV$ and $\alpha_{bs} = 0.67$. Assuming $\alpha_{bs} = 0$ the best-fit shifts to $\Lambda_{bs} \approx 38.5 \TeV$, corresponding to $\Delta C_9^\mu = - \Delta C_{10}^\mu \approx - 0.40$. We also note that the difference in $\chi^2$ between these two points is completely negligible. Indeed, the fit presents an approximate flat direction in $\alpha_{bs}$ for approximately $|\alpha_{bs}| \lesssim \pi/4$.

In case of the vector solution, $\Delta C_9^\mu$, the best-fit point assuming vanishing imaginary part is found for $\Delta C_9^\mu = -0.82$.

Lastly, a short comment is in order regarding the precision of this fit. It is well known that the cancellation of uncertainties in the ratios which define the clean observables is a feature that happens only for the SM point. When considering non-vanishing NP coefficients, the uncertainties in the knowledge of the form factors become relevant.
A precise fit should therefore include also these uncertainties and marginalise over the relevant parameters, this is however beyond the purpose of this work. Comparing the top-left panel in \Fig{bsmumuFit} with the analogous result of Ref.~\cite{Geng:2017svp} we check that our results are in good enough agreement with a more complete fit.

\section{Flavor observables} \label{App:Flavor observables}

We collect in this Appendix all relevant formulas for flavor observables
employed throughout this paper.

\subsection{$\text{Br}(B^{+}\to\pi^{+}\mu^{+}\mu^{-})_{\left[1,6\right]}$}

The structure of NP corrections to this observable is identical to
the one of $R_{K}$,
modulo a rescaling
$\frac{V_{td}^{*}}{V_{ts}^{*}}$ to the overall normalization of SM
Wilson coefficients. Neglecting the interference with the SM dipole operator, one has \cite{Hiller:2014ula,DAmico:2017mtc}

\be
\frac{\text{Br}(B^{+}\to\pi^{+}\mu^{+}\mu^{-})_{\left[1,6\right]}}{\text{Br}(B^{+}\to\pi^{+}\mu^{+}\mu^{-})_{\left[1,6\right]}^{\text{SM}}} \approx \frac{\left| C_{R}^{bd} -C_{L}^{bd} \right|^{2}+\left| C_{R}^{bd} + C_{L}^{bd} \right|^{2}}{\left| C_{R, \SM}^{bd} -C_{L, \SM}^{bd}\right|^{2}+\left| C_{R, \SM}^{bd} +C_{L, \SM}^{bd} \right|^{2} }~.
\label{eq:Br( B+->pi+ mu + mu- )}
\ee
For the SM prediction, we use the result from Ref. \cite{Khodjamirian:2017fxg}\footnote{We are not considering the lattice result also quoted in Ref. \cite{Khodjamirian:2017fxg},
as this result does not include non-local hadronic effects in the branching ratio.}:
\begin{equation}
\text{Br}(B^{+}\to\pi^{+}\mu^{+}\mu^{-})_{\left[1,6\right]}^{\text{SM}}=(6.55\pm1.25)\times10^{-9},\label{eq:Br(B+pi+mumu)_SM}
\end{equation}
whereas the experimentally measured value is \cite{Aaij:2015nea}:
\begin{equation}
\text{Br}(B^{+}\to\pi^{+}\mu^{+}\mu^{-})_{\left[1,6\right]}^{\text{exp}}=(4.55\pm1.05)\times10^{-9}.\label{eq:Br(B+pi+mumu)_exp}
\end{equation}

\subsection{$\text{Br}(B_{d}^{0}\to\mu^{+}\mu^{-})$}

In this case we consider Eq.~\eqref{eq:NPEFT}, with $ij=bd$.
The SM contribution to the operators is \cite{Buras:1998raa}
\begin{align}
C_{L,\text{SM}}^{bd} & =\frac{4G_{F}}{\sqrt{2}}\frac{\alpha(M_{Z})}{2\pi s_W ^2}V_{td}^{*}V_{tb}Y(x_t),\label{eq:L(bd mu mu)_SM}
\end{align}
while we neglect $\left|\frac{C_{R, \SM}^{bd}}{C_{L, \SM}^{bd}}\right|\approx2\%$. We obtain:
\begin{equation}
	\Gamma(B_{d}^{0}\to\mu^{+}\mu^{-})=\frac{\left| C_{R}^{bd} -C_{L}^{bd} \right|^{2}f_{B_{d}}^{2}m_{B_{d}}m_{\mu}^{2}}{32\pi}\sqrt{1-\frac{4m_{\mu}^{2}}{m_{B_{d}}^{2}}},
	\label{eq:Gamma(Bd mu mu)}
\end{equation}
where $C^{bd}_{L(R)} = C^{bd}_{L(R),\SM} + C^{bd}_{L(R), \text{NP}}$.
The strongest experimental constraints to date has been obtained by the ATLAS experiment \cite{Aaboud:2018mst}:
\begin{equation}
	\Br(B_{d}^{0}\to\mu^{+}\mu^{-})<2.1\times10^{-10}\ \text{(95\%\,CL)}~.
	\label{eq:Br(Bd mumu)_exp}
\end{equation}

\subsection{$\text{Br}(K_{L}\to\mu^{+}\mu^{-})_{\text{SD}}$ and $\text{Br}(K_{S}\to\mu^{+}\mu^{-})$}

The relevant effective lagrangian is given in Eq.~\eqref{eq:NPEFT}, with $ij=sd,ds$.
The Standard Model contribution to the two local operators is given by \cite{Buchalla:1993wq}
\be
 C_{L, \SM}^{sd} = \frac{4G_{F}}{\sqrt{2}}\frac{\alpha(M_{Z})}{2\pi s_{W}^{2}}(V_{cs}^{*}V_{cd}Y_{NL}+V_{ts}^{*}V_{td}Y(x_{t}))~, \qquad 
 C_{R, \SM}^{sd} = 0,
 \label{eq:L(sd mu mu)_SM}
\ee
where (neglecting theoretical uncertainties) $Y(x_t)=0.94$ and $Y_{NL}= 3\times 10^{-4}$.

Neglecting indirect CP violation (i.e. identifying $\vert K_{L}\rangle$
with the $CP$-odd combination of $\vert K^{0}\rangle$ and $\vert\overline{K}^{0}\rangle$)
and assuming $f_{K^{0}}=f_{K^{+}}$, we obtain:
\begin{align}
	\Gamma(K_{L}\to\mu^{+}\mu^{-})_{\text{SD}} & =\frac{\left[\text{Re}(C^{sd}_R -C_L^{sd})\right]^{2}f_{K}^{2}m_{K^{0}}m_{\mu}^{2}}{16\pi}\sqrt{1-\frac{4m_{\mu}^{2}}{m_{K^{0}}^{2}}},
	\label{eq:Gamma(KLmumu)_SD}
\end{align}
where $C^{sd}_{L(R)} = C^{sd}_{L(R),\SM} + C^{sd}_{L(R), \text{NP}}$.

In the case of $\Gamma(K_{S}\to\mu^{+}\mu^{-})$ the SM contribution, which is dominated by long distance physics, is negligible with respect to the (present and prospected) experimental bound, so that we can use:

\begin{align}
\Gamma(K_{S}\to\mu^{+}\mu^{-}) & =\frac{\left[\text{Im}(C^{sd}_R -C_L^{sd})\right]^{2}f_{K}^{2}m_{K^{0}}m_{\mu}^{2}}{16\pi}\sqrt{1-\frac{4m_{\mu}^{2}}{m_{K^{0}}^{2}}}.\label{eq:Gamma(KSmumu)}
\end{align}
The present limits are \cite{Ambrose:2000gj,Isidori:2003ts,Aaij:2017tia,LHCb:2019aoh}:
\begin{align}
\text{Br}(K_{L}\to\mu^{+}\mu^{-})_{\text{SD}} & <2.5\times10^{-9}\ ,\label{eq:Br(KLmumu)_SD_exp}\\
\text{Br}(K_{S}\to\mu^{+}\mu^{-}) & < 2.4 \times10^{-10}\ \text{(95\% CL)}~.\label{eq:Br(KSmumu)_exp}
\end{align}

\subsection{$\text{Br}(K_{L}\to \pi^0\mu^{+}\mu^{-})$}

The branching ratio in the SM can be calculated with good accuracy \cite{DAmbrosio:1998gur,Buchalla:2003sj,Isidori:2004rb,Mescia:2006jd,Bobeth:2017ecx}. Including NP contributions as in Eq.~\eqref{eq:SMEFTlagrangian} one has
\be
	\Br(K_L \to \pi^0 \mu^+ \mu^-) = (C^\mu_{\rm dir} \pm C^\mu_{\rm int} |a_S| + C^\mu_{\rm mix} |a_S|^2 C^\mu_{\gamma\gamma}) \times 10^{-12}~,
\ee
where $|a_S| \approx 1.2$, $C^\mu_{\gamma\gamma} \approx 5.2$, $C^\mu_{\rm mix} \approx 3.36$,
\be
	C^\mu_{\rm dir} \approx 1.09 (\omega_{7V}^2 + \omega_{7A}^2)~, \qquad
	C^\mu_{\rm int} \approx 2.63 \omega_{7V}~,
\ee
and the dependence on the coefficients of the operators in Eq.~\eqref{eq:NPEFT} is given by
\be\begin{split}
	\omega_{7V} &\approx 0.73 + \frac{1}{\Im(V_{ts}^* V_{td})} \Im \left( \frac{C^{sd}_R + C^{sd}_L}{2\sqrt{2}G_F \alpha} \right) \\
	\omega_{7A} &\approx -0.68 + \frac{1}{\Im(V_{ts}^* V_{td})} \Im \left( \frac{C^{sd}_R - C^{sd}_L}{2\sqrt{2}G_F \alpha} \right)~.
\end{split}\ee
The present experimental bound from KTeV \cite{AlaviHarati:2000hs} is
\be
	\Br(K_L \to \pi^0 \mu^+ \mu^-) < 3.8 \times 10^{-10} ~~~ (90\% \, {\rm CL}).
\ee

\subsection{$\text{Br}(K^{+}\to\pi^{+}\nu\overline{\nu})$ and $\text{Br}(K_{L}\to\pi^{0}\nu\overline{\nu})$}

The relevant effective lagrangian is: 
\begin{equation}
\mathcal{L}_{\text{eff}}^{sd\nu\nu}=(\overline{s}\gamma_{\rho}P_{L}d)\sum_{\ell}C^{sd,\ell}(\overline{\nu}_{\ell}\gamma^{\rho}P_{L}\nu_{\ell})+\text{h.c.},\label{eq:L_eff (sdnunu)}
\end{equation}
where \cite{Buras:1998raa}\footnote{The value of $X_{c}^{\ell}$ is interpolated from Table 29 of \cite[13.2.1]{Buras:1998raa}.}:

\begin{align}
C_{\text{SM}}^{sd,\ell} & =\left(\frac{4G_{F}}{\sqrt{2}}\frac{\alpha}{2\pi}V_{ts}^{*}V_{td}\right)c_{\ell}~,\label{eq:C^sdnunu_SM}\\
c_{\ell} & =-\frac{1}{s_{W}^{2}}\left(X_{t}+\frac{V_{cs}^{*}V_{cd}}{V_{ts}^{*}V_{td}}X_{c}^{\ell}\right)~,\label{eq:c_ell}\\
X_{t} & =1.481\pm0.009~,\label{eq:Xt}\\
X_{c}^{e} & =X_{c}^{\mu}=1.053\times10^{-3}~,\label{eq:X^e_c}\\
X_{c}^{\tau} & =0.711\times10^{-3}~.\label{eq:X^tau_c}
\end{align}
The corresponding formulae for branching ratios are:

\begin{align}
\text{Br}(K^{+}\to\pi^{+}\nu\nu) & =\text{Br}(K^{+}\to\pi^{+}\nu_{\tau}\overline{\nu}_{\tau})_{\text{SM}}+\text{Br}(K^{+}\to\pi^{+}\nu_{e}\overline{\nu}_{e})_{\text{SM}}\left\{ 1+\left|1+\frac{C_{\text{NP}}^{sd,\mu}}{C_{\text{SM}}^{sd,\mu}}\right|^{2}\right\} ,\label{eq:Br(K+->pi+nunu)}\\
\dfrac{\text{Br}(K_{L}\to\pi^{0}\nu\nu)}{\text{Br}(K_{L}\to\pi^{0}\nu\nu)_\text{SM}} & =\frac{2}{3}+\frac{1}{3}(1+\frac{\text{Im}C_{\text{NP}}^{sd,\mu}}{\text{Im}C_{\text{SM}}^{sd,\mu}})^{2}\label{eq:Br(K_L -> pi0 nu nu)}~.
\end{align}
The SM expressions of branching fractions can be found in \cite{Buras:1998raa}.
Using the values \eqref{eq:c_ell}-\eqref{eq:X^tau_c}, we find:
\begin{align}
\text{Br}(K^{+}\to\pi^{+}\nu_{e}\overline{\nu}_{e})_{\text{SM}} & =3.06\times10^{-11},\label{eq:Br(K+ -> pi+ nu_e nu_e)_SM}\\
\text{Br}(K^{+}\to\pi^{+}\nu_{\tau}\overline{\nu}_{\tau})_{\text{SM}} & =2.52\times10^{-11},\label{eq:Br(K+ -> pi+ nu_tau nu_tau)_SM}\\
\text{Br}(K_{L}\to\pi^{0}\nu\overline{\nu})_{\text{SM}} & =3.4\times10^{-11},\label{eq:Br(K_L->pi0nu nu)_SM}
\end{align}
where the theoretical uncertainty is negligible in comparison to the
experimental one.

The current experimental bounds are \cite{NA62:2019ajt,Artamonov:2008qb,Ahn:2018mvc,Tanabashi:2018oca}:\footnote{For $K^+ \to \pi^+ \nu \bar \nu$ we use the latest NA62 result from a combination of 2016 and 2017 data, presented at Kaon 2019, Perugia (\url{https://indico.cern.ch/event/769729/contributions/3510938/attachments/1905346/3146619/kaon2019_ruggiero_final.pdf}).}
\begin{align}
\text{Br}(K^{+}\to\pi^{+}\nu\nu)_{\text{exp}} &< 2.44 \times 10^{-10} \, @\, 95\% {\rm CL} \label{eq:Br(K+ -> pi+ nu_tau nu_tau)_exp}~,\\
\text{Br}(K_{L}\to\pi^{0}\nu\nu)_{\text{exp}} & <3.0\times10^{-9}  \, @\, 95\% {\rm CL} \label{eq:Br(K_L ->pi0 nu nu)_exp}~.
\end{align}

\subsection{Constraints on $\Delta F = 2$ operators}
\label{sec:DeltaF2limits}

\begin{table}[t]
\centering
\begin{tabular}{|c|}
\hline 
	Limits on $\Delta F = 2$ coefficients [\!$\GeV^{-2}$] \\ \hline \hline
	Re$C^1_K\in [-6.8, 7.7] \times 10^{-13}$~, Im$C^1_K \in [-1.2, 2.4] \times 10^{-15}$\\
	Re$C^1_D\in [-2.5, 3.1] \times 10^{-13}$~, Im$C^1_D \in [-9.4, 8.9] \times 10^{-15}$\\
	$ |C^1_{B_d}| < 9.5 \times 10^{-13} $ \\
	$ |C^1_{B_s}| < 1.9 \times 10^{-11} $ \\\hline
\end{tabular}
\caption{\label{tab:DeltaF2limits}Limits on $\Delta F = 2$ operators from meson anti-meson mixing from \cite{Bona:2007vi} (2018 update).}
\end{table}

In order to put limits on our simplified models from meson-antimeson mixing we use the results of Ref.~\cite{Bona:2007vi}, in particular the update presented at 'La Thuile 2018' by L. Silvestrini\footnote{The slides can be found at \url{https://agenda.infn.it/event/14377/contributions/24434/attachments/17481/19830/silvestriniLaThuile.pdf}.}. The relevant 95\%CL bounds, in $\GeV^{-2}$, on the coefficients $\Delta F = 2$ operators $(\bar q^i_L \gamma_\mu q^j_L)^2$, are summarised in Table~\ref{tab:DeltaF2limits}.
In case of the $S_3$ leptoquark, using Eq.~\eqref{eq:DeltaF2_S3} we calculate the maximum value of the leptoquark mass for any point in parameter space and show the result as dashed purple contours in Fig.~\ref{fig:S3limits}.
Similarly, in case of the $Z^\prime$, we set upper an limit on its mass, assuming a maximum value for the $g_\mu$ coupling from perturbative unitarity, c.f. Sec.~\ref{sec:Zprime}.

Using the dependence of $\epsilon^\prime / \epsilon$ on the coefficients of four-quark operators of the type $(\bar s \gamma_\mu P_L d)(\bar q \gamma^\mu P_L q)$ from Refs.~\cite{Aebischer:2018quc,Aebischer:2018csl} we checked that the constraint this observable, in our framework, is not competitive with those from meson-antimeson mixing.


{\small
\bibliography{Biblio}{}

\providecommand{\href}[2]{#2}\begingroup\raggedright\begin{thebibliography}{10}

\bibitem{Aaij:2014ora}
{\bf LHCb} Collaboration, R.~Aaij et~al. {\em Phys. Rev. Lett.} {\bf 113}
  (2014) 151601, [\href{http://arxiv.org/abs/1406.6482}{{\tt
  arXiv:1406.6482}}].

\bibitem{Aaij:2019aa}
{\bf LHCb} Collaboration, R.~Aaij et~al.
  \href{http://arxiv.org/abs/1903.09252}{{\tt arXiv:1903.09252}}.

\bibitem{Aaij:2017vbb}
{\bf LHCb} Collaboration, R.~Aaij et~al. {\em JHEP} {\bf 08} (2017) 055,
  [\href{http://arxiv.org/abs/1705.05802}{{\tt arXiv:1705.05802}}].

\bibitem{2019:BelleRKst}
{\bf Belle} Collaboration, {\it Seminar at rencontres de moriond ew},  2019.

\bibitem{Aaij:2014pli}
{\bf LHCb} Collaboration, R.~Aaij et~al. {\em JHEP} {\bf 06} (2014) 133,
  [\href{http://arxiv.org/abs/1403.8044}{{\tt arXiv:1403.8044}}].

\bibitem{Aaij:2015esa}
{\bf LHCb} Collaboration, R.~Aaij et~al. {\em JHEP} {\bf 09} (2015) 179,
  [\href{http://arxiv.org/abs/1506.08777}{{\tt arXiv:1506.08777}}].

\bibitem{Aaij:2013qta}
{\bf LHCb} Collaboration, R.~Aaij et~al. {\em Phys. Rev. Lett.} {\bf 111}
  (2013) 191801, [\href{http://arxiv.org/abs/1308.1707}{{\tt
  arXiv:1308.1707}}].

\bibitem{Aaij:2015oid}
{\bf LHCb} Collaboration, R.~Aaij et~al. {\em JHEP} {\bf 02} (2016) 104,
  [\href{http://arxiv.org/abs/1512.04442}{{\tt arXiv:1512.04442}}].

\bibitem{Aaij:2017vad}
{\bf LHCb} Collaboration, R.~Aaij et~al. {\em Phys. Rev. Lett.} {\bf 118}
  (2017), no.~19 191801, [\href{http://arxiv.org/abs/1703.05747}{{\tt
  arXiv:1703.05747}}].

\bibitem{Aaboud:2018mst}
{\bf ATLAS} Collaboration, M.~Aaboud et~al. {\em Submitted to: JHEP} (2018)
  [\href{http://arxiv.org/abs/1812.03017}{{\tt arXiv:1812.03017}}].

\bibitem{Hiller:2014yaa}
G.~Hiller and M.~Schmaltz {\em Phys. Rev.} {\bf D90} (2014) 054014,
  [\href{http://arxiv.org/abs/1408.1627}{{\tt arXiv:1408.1627}}].

\bibitem{Descotes-Genon:2015uva}
S.~Descotes-Genon, L.~Hofer, J.~Matias, and J.~Virto {\em JHEP} {\bf 06} (2016)
  092, [\href{http://arxiv.org/abs/1510.04239}{{\tt arXiv:1510.04239}}].

\bibitem{Altmannshofer:2017fio}
W.~Altmannshofer, C.~Niehoff, P.~Stangl, and D.~M. Straub {\em Eur. Phys. J.}
  {\bf C77} (2017), no.~6 377, [\href{http://arxiv.org/abs/1703.09189}{{\tt
  arXiv:1703.09189}}].

\bibitem{Capdevila:2017bsm}
B.~Capdevila, A.~Crivellin, S.~Descotes-Genon, J.~Matias, and J.~Virto {\em
  JHEP} {\bf 01} (2018) 093, [\href{http://arxiv.org/abs/1704.05340}{{\tt
  arXiv:1704.05340}}].

\bibitem{DAmico:2017mtc}
G.~D'Amico, M.~Nardecchia, P.~Panci, F.~Sannino, A.~Strumia, R.~Torre, and
  A.~Urbano {\em JHEP} {\bf 09} (2017) 010,
  [\href{http://arxiv.org/abs/1704.05438}{{\tt arXiv:1704.05438}}].

\bibitem{Altmannshofer:2017yso}
W.~Altmannshofer, P.~Stangl, and D.~M. Straub {\em Phys. Rev.} {\bf D96}
  (2017), no.~5 055008, [\href{http://arxiv.org/abs/1704.05435}{{\tt
  arXiv:1704.05435}}].

\bibitem{Geng:2017svp}
L.-S. Geng, B.~Grinstein, S.~J{\"a}ger, J.~Martin~Camalich, X.-L. Ren, and
  R.-X. Shi {\em Phys. Rev.} {\bf D96} (2017), no.~9 093006,
  [\href{http://arxiv.org/abs/1704.05446}{{\tt arXiv:1704.05446}}].

\bibitem{Ciuchini:2017mik}
M.~Ciuchini, A.~M. Coutinho, M.~Fedele, E.~Franco, A.~Paul, L.~Silvestrini, and
  M.~Valli {\em Eur. Phys. J.} {\bf C77} (2017), no.~10 688,
  [\href{http://arxiv.org/abs/1704.05447}{{\tt arXiv:1704.05447}}].

\bibitem{Hiller:2017bzc}
G.~Hiller and I.~Nisandzic {\em Phys. Rev.} {\bf D96} (2017), no.~3 035003,
  [\href{http://arxiv.org/abs/1704.05444}{{\tt arXiv:1704.05444}}].

\bibitem{Alok:2017sui}
A.~K. Alok, B.~Bhattacharya, A.~Datta, D.~Kumar, J.~Kumar, and D.~London {\em
  Phys. Rev.} {\bf D96} (2017), no.~9 095009,
  [\href{http://arxiv.org/abs/1704.07397}{{\tt arXiv:1704.07397}}].

\bibitem{Hurth:2017hxg}
T.~Hurth, F.~Mahmoudi, D.~Martinez~Santos, and S.~Neshatpour {\em Phys. Rev.}
  {\bf D96} (2017), no.~9 095034, [\href{http://arxiv.org/abs/1705.06274}{{\tt
  arXiv:1705.06274}}].

\bibitem{Alguero:2019aa}
M.~Alguer{\'o}, B.~Capdevila, A.~Crivellin, S.~Descotes-Genon, P.~Masjuan,
  J.~Matias, and J.~Virto \href{http://arxiv.org/abs/1903.09578}{{\tt
  arXiv:1903.09578}}.

\bibitem{Alok:2019ufo}
A.~K. Alok, A.~Dighe, S.~Gangal, and D.~Kumar
  \href{http://arxiv.org/abs/1903.09617}{{\tt arXiv:1903.09617}}.

\bibitem{Ciuchini:2019usw}
M.~Ciuchini, A.~M. Coutinho, M.~Fedele, E.~Franco, A.~Paul, L.~Silvestrini, and
  M.~Valli \href{http://arxiv.org/abs/1903.09632}{{\tt arXiv:1903.09632}}.

\bibitem{Aebischer:2019mlg}
J.~Aebischer, W.~Altmannshofer, D.~Guadagnoli, M.~Reboud, P.~Stangl, and D.~M.
  Straub \href{http://arxiv.org/abs/1903.10434}{{\tt arXiv:1903.10434}}.

\bibitem{Alguero:2018nvb}
M.~Alguer{\'o}, B.~Capdevila, S.~Descotes-Genon, P.~Masjuan, and J.~Matias
  \href{http://arxiv.org/abs/1809.08447}{{\tt arXiv:1809.08447}}.

\bibitem{Alguero:2019pjc}
M.~Alguer{\'o}, B.~Capdevila, S.~Descotes-Genon, P.~Masjuan, and J.~Matias
  \href{http://arxiv.org/abs/1902.04900}{{\tt arXiv:1902.04900}}.

\bibitem{Datta:2019zca}
A.~Datta, J.~Kumar, and D.~London \href{http://arxiv.org/abs/1903.10086}{{\tt
  arXiv:1903.10086}}.

\bibitem{DAmbrosio:2002vsn}
G.~D'Ambrosio, G.~F. Giudice, G.~Isidori, and A.~Strumia {\em Nucl. Phys.} {\bf
  B645} (2002) 155--187, [\href{http://arxiv.org/abs/hep-ph/0207036}{{\tt
  hep-ph/0207036}}].

\bibitem{Barbieri:2011ci}
R.~Barbieri, G.~Isidori, J.~Jones-Perez, P.~Lodone, and D.~M. Straub {\em Eur.
  Phys. J.} {\bf C71} (2011) 1725, [\href{http://arxiv.org/abs/1105.2296}{{\tt
  arXiv:1105.2296}}].

\bibitem{Barbieri:2012uh}
R.~Barbieri, D.~Buttazzo, F.~Sala, and D.~M. Straub {\em JHEP} {\bf 07} (2012)
  181, [\href{http://arxiv.org/abs/1203.4218}{{\tt arXiv:1203.4218}}].

\bibitem{Greljo:2015mma}
A.~Greljo, G.~Isidori, and D.~Marzocca {\em JHEP} {\bf 07} (2015) 142,
  [\href{http://arxiv.org/abs/1506.01705}{{\tt arXiv:1506.01705}}].

\bibitem{Barbieri:2015yvd}
R.~Barbieri, G.~Isidori, A.~Pattori, and F.~Senia {\em Eur. Phys. J.} {\bf C76}
  (2016), no.~2 67, [\href{http://arxiv.org/abs/1512.01560}{{\tt
  arXiv:1512.01560}}].

\bibitem{Barbieri:2016las}
R.~Barbieri, C.~W. Murphy, and F.~Senia {\em Eur. Phys. J.} {\bf C77} (2017),
  no.~1 8, [\href{http://arxiv.org/abs/1611.04930}{{\tt arXiv:1611.04930}}].

\bibitem{Bordone:2017anc}
M.~Bordone, G.~Isidori, and S.~Trifinopoulos {\em Phys. Rev.} {\bf D96} (2017),
  no.~1 015038, [\href{http://arxiv.org/abs/1702.07238}{{\tt
  arXiv:1702.07238}}].

\bibitem{Bordone:2017lsy}
M.~Bordone, D.~Buttazzo, G.~Isidori, and J.~Monnard {\em Eur. Phys. J.} {\bf
  C77} (2017), no.~9 618, [\href{http://arxiv.org/abs/1705.10729}{{\tt
  arXiv:1705.10729}}].

\bibitem{Buttazzo:2017ixm}
D.~Buttazzo, A.~Greljo, G.~Isidori, and D.~Marzocca {\em JHEP} {\bf 11} (2017)
  044, [\href{http://arxiv.org/abs/1706.07808}{{\tt arXiv:1706.07808}}].

\bibitem{Barbieri:2017tuq}
R.~Barbieri and A.~Tesi {\em Eur. Phys. J.} {\bf C78} (2018), no.~3 193,
  [\href{http://arxiv.org/abs/1712.06844}{{\tt arXiv:1712.06844}}].

\bibitem{Glashow:2014iga}
S.~L. Glashow, D.~Guadagnoli, and K.~Lane {\em Phys. Rev. Lett.} {\bf 114}
  (2015) 091801, [\href{http://arxiv.org/abs/1411.0565}{{\tt
  arXiv:1411.0565}}].

\bibitem{Bhattacharya:2014wla}
B.~Bhattacharya, A.~Datta, D.~London, and S.~Shivashankara {\em Phys. Lett.}
  {\bf B742} (2015) 370--374, [\href{http://arxiv.org/abs/1412.7164}{{\tt
  arXiv:1412.7164}}].

\bibitem{Angelescu:2018tyl}
A.~Angelescu, D.~Be{\v c}irevi{\'c}, D.~A. Faroughy, and O.~Sumensari {\em
  JHEP} {\bf 10} (2018) 183, [\href{http://arxiv.org/abs/1808.08179}{{\tt
  arXiv:1808.08179}}].

\bibitem{Gripaios:2015gra}
B.~Gripaios, M.~Nardecchia, and S.~A. Renner {\em JHEP} {\bf 06} (2016) 083,
  [\href{http://arxiv.org/abs/1509.05020}{{\tt arXiv:1509.05020}}].

\bibitem{Cline:2017aed}
J.~M. Cline {\em Phys. Rev.} {\bf D97} (2018), no.~1 015013,
  [\href{http://arxiv.org/abs/1710.02140}{{\tt arXiv:1710.02140}}].

\bibitem{Cline:2017qqu}
J.~M. Cline and J.~M. Cornell {\em Phys. Lett.} {\bf B782} (2018) 232--237,
  [\href{http://arxiv.org/abs/1711.10770}{{\tt arXiv:1711.10770}}].

\bibitem{Bobeth:2013uxa}
C.~Bobeth, M.~Gorbahn, T.~Hermann, M.~Misiak, E.~Stamou, and M.~Steinhauser
  {\em Phys. Rev. Lett.} {\bf 112} (2014) 101801,
  [\href{http://arxiv.org/abs/1311.0903}{{\tt arXiv:1311.0903}}].

\bibitem{Aaij:2015nea}
{\bf LHCb} Collaboration, R.~Aaij et~al. {\em JHEP} {\bf 10} (2015) 034,
  [\href{http://arxiv.org/abs/1509.00414}{{\tt arXiv:1509.00414}}].

\bibitem{Du:2015tda}
D.~Du, A.~X. El-Khadra, S.~Gottlieb, A.~S. Kronfeld, J.~Laiho, E.~Lunghi, R.~S.
  Van~de Water, and R.~Zhou {\em Phys. Rev.} {\bf D93} (2016), no.~3 034005,
  [\href{http://arxiv.org/abs/1510.02349}{{\tt arXiv:1510.02349}}].

\bibitem{Khodjamirian:2017fxg}
A.~Khodjamirian and A.~V. Rusov {\em JHEP} {\bf 08} (2017) 112,
  [\href{http://arxiv.org/abs/1703.04765}{{\tt arXiv:1703.04765}}].

\bibitem{Aaij:2017tia}
{\bf LHCb} Collaboration, R.~Aaij et~al. {\em Eur. Phys. J.} {\bf C77} (2017),
  no.~10 678, [\href{http://arxiv.org/abs/1706.00758}{{\tt arXiv:1706.00758}}].

\bibitem{LHCb:2019aoh}
{\bf LHCb} Collaboration, {\it {Strong constraints on the $K^0_s \to \mu^+
  \mu^-$ branching fraction}},  No.~LHCb-CONF-2019-002, 2019.

\bibitem{Ambrose:2000gj}
{\bf E871} Collaboration, D.~Ambrose et~al. {\em Phys. Rev. Lett.} {\bf 84}
  (2000) 1389--1392.

\bibitem{Isidori:2003ts}
G.~Isidori and R.~Unterdorfer {\em JHEP} {\bf 01} (2004) 009,
  [\href{http://arxiv.org/abs/hep-ph/0311084}{{\tt hep-ph/0311084}}].

\bibitem{AlaviHarati:2000hs}
{\bf KTEV} Collaboration, A.~Alavi-Harati et~al. {\em Phys. Rev. Lett.} {\bf
  84} (2000) 5279--5282, [\href{http://arxiv.org/abs/hep-ex/0001006}{{\tt
  hep-ex/0001006}}].

\bibitem{DAmbrosio:1998gur}
G.~D'Ambrosio, G.~Ecker, G.~Isidori, and J.~Portoles {\em JHEP} {\bf 08} (1998)
  004, [\href{http://arxiv.org/abs/hep-ph/9808289}{{\tt hep-ph/9808289}}].

\bibitem{Buchalla:2003sj}
G.~Buchalla, G.~D'Ambrosio, and G.~Isidori {\em Nucl. Phys.} {\bf B672} (2003)
  387--408, [\href{http://arxiv.org/abs/hep-ph/0308008}{{\tt hep-ph/0308008}}].

\bibitem{Isidori:2004rb}
G.~Isidori, C.~Smith, and R.~Unterdorfer {\em Eur. Phys. J.} {\bf C36} (2004)
  57--66, [\href{http://arxiv.org/abs/hep-ph/0404127}{{\tt hep-ph/0404127}}].

\bibitem{Mescia:2006jd}
F.~Mescia, C.~Smith, and S.~Trine {\em JHEP} {\bf 08} (2006) 088,
  [\href{http://arxiv.org/abs/hep-ph/0606081}{{\tt hep-ph/0606081}}].

\bibitem{Aaij:2012rt}
{\bf LHCb} Collaboration, R.~Aaij et~al. {\em JHEP} {\bf 01} (2013) 090,
  [\href{http://arxiv.org/abs/1209.4029}{{\tt arXiv:1209.4029}}].

\bibitem{Bobeth:2017ecx}
C.~Bobeth and A.~J. Buras {\em JHEP} {\bf 02} (2018) 101,
  [\href{http://arxiv.org/abs/1712.01295}{{\tt arXiv:1712.01295}}].

\bibitem{NA62:2019ajt}
{\bf NA62} Collaboration, {\it {New Results on $K^+ \to \pi^+ \nu \bar\nu$ from
  the NA62 Experiment}},  Kaon 2019, Perugia, 2019.

\bibitem{Buras:2015qea}
A.~J. Buras, D.~Buttazzo, J.~Girrbach-Noe, and R.~Knegjens {\em JHEP} {\bf 11}
  (2015) 033, [\href{http://arxiv.org/abs/1503.02693}{{\tt arXiv:1503.02693}}].

\bibitem{Ahn:2018mvc}
{\bf KOTO} Collaboration, J.~K. Ahn et~al. {\em Phys. Rev. Lett.} {\bf 122}
  (2019), no.~2 021802, [\href{http://arxiv.org/abs/1810.09655}{{\tt
  arXiv:1810.09655}}].

\bibitem{Aaboud:2017buh}
{\bf ATLAS} Collaboration, M.~Aaboud et~al. {\em JHEP} {\bf 10} (2017) 182,
  [\href{http://arxiv.org/abs/1707.02424}{{\tt arXiv:1707.02424}}].

\bibitem{Greljo:2017vvb}
A.~Greljo and D.~Marzocca {\em Eur. Phys. J.} {\bf C77} (2017), no.~8 548,
  [\href{http://arxiv.org/abs/1704.09015}{{\tt arXiv:1704.09015}}].

\bibitem{Gripaios:2014tna}
B.~Gripaios, M.~Nardecchia, and S.~A. Renner {\em JHEP} {\bf 05} (2015) 006,
  [\href{http://arxiv.org/abs/1412.1791}{{\tt arXiv:1412.1791}}].

\bibitem{Varzielas:2015iva}
I.~de~Medeiros~Varzielas and G.~Hiller {\em JHEP} {\bf 06} (2015) 072,
  [\href{http://arxiv.org/abs/1503.01084}{{\tt arXiv:1503.01084}}].

\bibitem{Dorsner:2017ufx}
I.~Dor{\v s}ner, S.~Fajfer, D.~A. Faroughy, and N.~Ko{\v s}nik
  \href{http://arxiv.org/abs/1706.07779}{{\tt arXiv:1706.07779}}.
  [JHEP10,188(2017)].

\bibitem{Alonso:2015sja}
R.~Alonso, B.~Grinstein, and J.~Martin~Camalich {\em JHEP} {\bf 10} (2015) 184,
  [\href{http://arxiv.org/abs/1505.05164}{{\tt arXiv:1505.05164}}].

\bibitem{DiLuzio:2017vat}
L.~Di~Luzio, A.~Greljo, and M.~Nardecchia {\em Phys. Rev.} {\bf D96} (2017),
  no.~11 115011, [\href{http://arxiv.org/abs/1708.08450}{{\tt
  arXiv:1708.08450}}].

\bibitem{Bordone:2017bld}
M.~Bordone, C.~Cornella, J.~Fuentes-Martin, and G.~Isidori {\em Phys. Lett.}
  {\bf B779} (2018) 317--323, [\href{http://arxiv.org/abs/1712.01368}{{\tt
  arXiv:1712.01368}}].

\bibitem{DiLuzio:2018zxy}
L.~Di~Luzio, J.~Fuentes-Martin, A.~Greljo, M.~Nardecchia, and S.~Renner {\em
  JHEP} {\bf 11} (2018) 081, [\href{http://arxiv.org/abs/1808.00942}{{\tt
  arXiv:1808.00942}}].

\bibitem{Bordone:2018nbg}
M.~Bordone, C.~Cornella, J.~Fuentes-Mart{\'\i}n, and G.~Isidori {\em JHEP} {\bf
  10} (2018) 148, [\href{http://arxiv.org/abs/1805.09328}{{\tt
  arXiv:1805.09328}}].

\bibitem{Crivellin:2018yvo}
A.~Crivellin, C.~Greub, D.~M{\"u}ller, and F.~Saturnino {\em Phys. Rev. Lett.}
  {\bf 122} (2019), no.~1 011805, [\href{http://arxiv.org/abs/1807.02068}{{\tt
  arXiv:1807.02068}}].

\bibitem{Baker:2019sli}
M.~J. Baker, J.~Fuentes-Mart{\'\i}n, G.~Isidori, and M.~K{\"o}nig
  \href{http://arxiv.org/abs/1901.10480}{{\tt arXiv:1901.10480}}.

\bibitem{Altmannshofer:2014cfa}
W.~Altmannshofer, S.~Gori, M.~Pospelov, and I.~Yavin {\em Phys. Rev.} {\bf D89}
  (2014) 095033, [\href{http://arxiv.org/abs/1403.1269}{{\tt
  arXiv:1403.1269}}].

\bibitem{Crivellin:2015mga}
A.~Crivellin, G.~D'Ambrosio, and J.~Heeck {\em Phys. Rev. Lett.} {\bf 114}
  (2015) 151801, [\href{http://arxiv.org/abs/1501.00993}{{\tt
  arXiv:1501.00993}}].

\bibitem{Bonilla:2017lsq}
C.~Bonilla, T.~Modak, R.~Srivastava, and J.~W.~F. Valle {\em Phys. Rev.} {\bf
  D98} (2018), no.~9 095002, [\href{http://arxiv.org/abs/1705.00915}{{\tt
  arXiv:1705.00915}}].

\bibitem{Biswas:2019twf}
A.~Biswas and A.~Shaw \href{http://arxiv.org/abs/1903.08745}{{\tt
  arXiv:1903.08745}}.

\bibitem{DiLuzio:2017chi}
L.~Di~Luzio and M.~Nardecchia {\em Eur. Phys. J.} {\bf C77} (2017), no.~8 536,
  [\href{http://arxiv.org/abs/1706.01868}{{\tt arXiv:1706.01868}}].

\bibitem{Bediaga:2018lhg}
{\bf LHCb} Collaboration, R.~Aaij et~al.
  \href{http://arxiv.org/abs/1808.08865}{{\tt arXiv:1808.08865}}.

\bibitem{Kou:2018nap}
{\bf Belle II} Collaboration, W.~Altmannshofer et~al.
  \href{http://arxiv.org/abs/1808.10567}{{\tt arXiv:1808.10567}}.

\bibitem{Cerri:2018ypt}
A.~Cerri et~al. \href{http://arxiv.org/abs/1812.07638}{{\tt arXiv:1812.07638}}.

\bibitem{Ambrosino:2019qvz}
{\bf KLEVER Project} Collaboration, F.~Ambrosino et~al.
  \href{http://arxiv.org/abs/1901.03099}{{\tt arXiv:1901.03099}}.

\bibitem{Ruggiero:2017hjh}
{\bf NA62} Collaboration, G.~Ruggiero {\em J. Phys. Conf. Ser.} {\bf 800}
  (2017), no.~1 012023.

\bibitem{Bordone:2016gaq}
M.~Bordone, G.~Isidori, and A.~Pattori {\em Eur. Phys. J.} {\bf C76} (2016),
  no.~8 440, [\href{http://arxiv.org/abs/1605.07633}{{\tt arXiv:1605.07633}}].

\bibitem{Altmannshofer:2008dz}
W.~Altmannshofer, P.~Ball, A.~Bharucha, A.~J. Buras, D.~M. Straub, and M.~Wick
  {\em JHEP} {\bf 01} (2009) 019, [\href{http://arxiv.org/abs/0811.1214}{{\tt
  arXiv:0811.1214}}].

\bibitem{Bobeth:2011nj}
C.~Bobeth, G.~Hiller, D.~van Dyk, and C.~Wacker {\em JHEP} {\bf 01} (2012) 107,
  [\href{http://arxiv.org/abs/1111.2558}{{\tt arXiv:1111.2558}}].

\bibitem{Becirevic:2012fy}
D.~Becirevic, N.~Kosnik, F.~Mescia, and E.~Schneider {\em Phys. Rev.} {\bf D86}
  (2012) 034034, [\href{http://arxiv.org/abs/1205.5811}{{\tt
  arXiv:1205.5811}}].

\bibitem{Straub:2015ica}
A.~Bharucha, D.~M. Straub, and R.~Zwicky {\em JHEP} {\bf 08} (2016) 098,
  [\href{http://arxiv.org/abs/1503.05534}{{\tt arXiv:1503.05534}}].

\bibitem{Altmannshofer:2014rta}
W.~Altmannshofer and D.~M. Straub {\em Eur. Phys. J.} {\bf C75} (2015), no.~8
  382, [\href{http://arxiv.org/abs/1411.3161}{{\tt arXiv:1411.3161}}].

\bibitem{Hiller:2014ula}
G.~Hiller and M.~Schmaltz {\em JHEP} {\bf 02} (2015) 055,
  [\href{http://arxiv.org/abs/1411.4773}{{\tt arXiv:1411.4773}}].

\bibitem{Buras:1998raa}
A.~J. Buras pp.~281--539, 1998.
\newblock \href{http://arxiv.org/abs/hep-ph/9806471}{{\tt hep-ph/9806471}}.

\bibitem{Buchalla:1993wq}
G.~Buchalla and A.~J. Buras {\em Nucl. Phys.} {\bf B412} (1994) 106--142,
  [\href{http://arxiv.org/abs/hep-ph/9308272}{{\tt hep-ph/9308272}}].

\bibitem{Artamonov:2008qb}
{\bf E949} Collaboration, A.~V. Artamonov et~al. {\em Phys. Rev. Lett.} {\bf
  101} (2008) 191802, [\href{http://arxiv.org/abs/0808.2459}{{\tt
  arXiv:0808.2459}}].

\bibitem{Tanabashi:2018oca}
{\bf Particle Data Group} Collaboration, M.~Tanabashi et~al. {\em Phys. Rev.}
  {\bf D98} (2018), no.~3 030001.

\bibitem{Bona:2007vi}
{\bf UTfit} Collaboration, M.~Bona et~al. {\em JHEP} {\bf 03} (2008) 049,
  [\href{http://arxiv.org/abs/0707.0636}{{\tt arXiv:0707.0636}}].

\bibitem{Aebischer:2018quc}
J.~Aebischer, C.~Bobeth, A.~J. Buras, J.-M. G{\'e}rard, and D.~M. Straub
  \href{http://arxiv.org/abs/1807.02520}{{\tt arXiv:1807.02520}}.

\bibitem{Aebischer:2018csl}
J.~Aebischer, C.~Bobeth, A.~J. Buras, and D.~M. Straub {\em Eur. Phys. J.} {\bf
  C79} (2019), no.~3 219, [\href{http://arxiv.org/abs/1808.00466}{{\tt
  arXiv:1808.00466}}].

\end{thebibliography}\endgroup
\bibliographystyle{JHEP}}

\end{document}